\documentclass[twocolumn,astrosymb]{aastex631}

\usepackage{amsmath}
\usepackage{natbib}
\defcitealias{pettini2004}{PP04}
\defcitealias{steidel2014}{Steidel14}
\defcitealias{strom2018}{Strom18}

\shorttitle{Chemical abundance patterns in $z\simeq2-3$ galaxies}
\shortauthors{Strom et~al.}
\graphicspath{{./}{figures/}}

\begin{document}

\title{Chemical abundance scaling relations for multiple elements in $z\simeq2-3$ star-forming galaxies}

\author[0000-0001-6369-1636]{Allison L. Strom}
\altaffiliation{Carnegie-Princeton Fellow}
\affiliation{Department of Astrophysical Sciences, 4 Ivy Lane, Princeton University, Princeton, NJ 08544, USA}
\affiliation{Carnegie Observatories, 813 Santa Barbara Street, Pasadena, CA 91101, USA}
\author[0000-0002-8459-5413]{Gwen C. Rudie}
\affiliation{Carnegie Observatories, 813 Santa Barbara Street, Pasadena, CA 91101, USA}
\author[0000-0002-4834-7260]{Charles C. Steidel}
\affiliation{Cahill Center for Astronomy and Astrophysics, California Institute of Technology, MS 249-17, Pasadena, CA 91125, USA}
\author[0000-0002-6967-7322]{Ryan F. Trainor}
\affiliation{Department of Physics and Astronomy, Franklin \& Marshall College, 637 College Avenue, Lancaster, PA 17603, USA}
\correspondingauthor{Allison L. Strom}
\email{allison.strom@princeton.edu}

\begin{abstract}
The chemical abundance patterns of gas and stars in galaxies are powerful probes of galaxies' star formation histories and the astrophysics of galaxy assembly but are challenging to measure with confidence in distant galaxies. In this paper, we report the first measurements of the correlation between stellar mass (M$_{\ast}$) and multiple tracers of chemical enrichment (including O, N, and Fe) in individual $z\sim2-3$ galaxies, using a sample of 195 star-forming galaxies from the Keck Baryonic Structure Survey (KBSS). The galaxies' chemical abundances are inferred using photoionization models capable of reconciling high-redshift galaxies' observed extreme rest-UV and rest-optical spectroscopic properties. We find that the M$_{\ast}$-O/H relation for our sample is relatively shallow, with moderately large scatter, and is offset $\sim0.35$~dex higher than the corresponding M$_{\ast}$-Fe/H relation. The two relations have very similar slopes, indicating a high level of $\alpha$-enhancement---$\textrm{O/Fe}\approx2.2\times(\textrm{O/Fe})_{\odot}$---across two decades in M$_{\ast}$. The M$_{\ast}$-N/H relation has the steepest slope and largest intrinsic scatter, which likely results from the fact that many $z\sim2$ galaxies are observed near or past the transition from ``primary" to ``secondary" N production and may reflect uncertainties in the astrophysical origin of N. Together, these results suggest that $z\sim2$ galaxies are old enough to have seen substantial enrichment from intermediate mass stars, but are still young enough that Type~Ia supernovae have not had time to contribute significantly to their enrichment.
\end{abstract}

\section{Introduction}
\label{sec:intro}

For over four decades, scaling relations pertaining to the chemical enrichment of galaxies' interstellar medium (ISM) have been used to study galaxy formation. \citet{lequeux1979} investigated the correlation between galaxy luminosity and gas-phase oxygen abundance and found that more luminous galaxies hosted \ion{H}{2} regions with higher O/H. Later, large surveys like the Sloan Digital Sky Survey \citep[SDSS;][]{york2000} enabled studies of large statistical samples of nearby galaxies. \citet{tremonti2004} were among the first to complete a detailed analysis and found a strong positive correlation between stellar mass (M$_{\ast}$) and gas-phase O/H, which was relatively steep at lower masses and started to flatten at M$_{\ast}\gtrsim10^{10.5}$~M$_{\odot}$. The M$_{\ast}$-O/H relation---often referred to as the ``mass-metallicity relation" or simply the MZR---has continued to be studied extensively at $z\sim0$ (see Section~5 of \citealt{maiolino2019} for a review) and has also been studied using resolved observations of nearby galaxies \citep[e.g.,][]{gonzalez-delgado2014,barrera-ballesteros2017,sanchez2019}. As the size and quality of high-redshift spectroscopic surveys has increased, the MZR has been confirmed to exist up to at least $z\sim3.5$ \citep[e.g.,][among many others]{erb2006metal,maiolino2008,steidel2014,zahid2014,onodera2016,sanders2021}.

Still, connecting the results from these studies and constructing a single clear picture of chemical enrichment over cosmic time has remained difficult, in large part because of the many different methods used to infer O/H. Although some studies of individual nearby objects have used ``direct method" abundances based on measurements of the electron temperature\footnote{Oxygen and other heavy elements provide the principal cooling mechanism in photoionized gas, linking temperature to enrichment.} ($T_e$) in \ion{H}{2} regions \citep[e.g.,][]{aller1954,pilyugin2005,berg2012,berg2020,perez-montero2014}, the spectral lines required to determine $T_e$ are exceedingly faint. Therefore, studies of faint, distant, and/or metal-rich galaxies have relied on more indirect tracers of chemical enrichment. The most common method is to use the ratios of strong emission lines in galaxy spectra, which can be empirically linked to gas-phase O/H using direct method abundances measured for $z\sim0$ samples \citep[e.g.,][]{mcgaugh1991,pettini2004}. Even at the same redshift as the calibration samples, however, the MZRs determined using different ``strong-line" methods often do not agree and differ substantially in both shape and normalization \citep{kewley2008,andrews2013}.

There are additional concerns about the application of strong-line methods at high redshift, where galaxies have significantly different stellar populations and conditions in their interstellar gas. The crux is that line ratio diagnostics are frequently sensitive to more than one parameter---e.g., O/H \emph{and} the abundance of other elements like nitrogen, or O/H \emph{and} the ionization conditions in the gas---many of which are observed to be correlated with one another in nearby galaxies. Indeed, the implicit dependence on these underlying correlations is fundamental to the success of strong-line diagnostics. Unfortunately, we do not know \textit{a priori} whether these properties are correlated with one another in the same way in high-$z$ galaxies as in $z\sim0$ galaxies, and so caution must be exercised when using locally-calibrated line ratio diagnostics to study distant galaxies.

Ideally, we would ``re-calibrate" common line ratio diagnostics using a representative sample of high-$z$ galaxies with measurements of $T_e$. However, despite heroic efforts from the ground \citep{yuan2009,erb2010,steidel2014,sanders2016,kojima2017,sanders2020}, this has not yet been achieved, because measurable $T_e$ corresponds to a relatively narrow range of O/H. Over the next several years, we can hope to revisit the task of constructing new, $T_e$-calibrated diagnostics that are more appropriate for measuring O/H at high redshift using observations with the James Webb Space Telescope \citep[JWST;][]{gardner2006}, scheduled to be launched later this year.

Even so, oxygen abundance is only part of the picture. In the nearby universe, it is common to study the chemical enrichment of galaxies using multiple elements. These studies frequently use observations of the stellar continuum to measure, e.g., the iron, magnesium, and carbon content of the integrated stellar population; see Section~2 of \citet{maiolino2019} for a description of the methodology, but also \citet{gallazzi2005}, which is the complementary study of stellar metallicity in the same galaxies studied by \citet{tremonti2004}. More detailed abundance patterns are also central to studies of the circumgalactic medium (CGM) at all redshifts \citep{tumlinson2017,zahedy2019,zahedy2021}. More recently, very deep spectra of massive, quiescent galaxies at high redshift have enabled the measurement of multiple elemental abundances \citep{kriek2016,kriek2019,jafariyazani2020}. Using deep rest-UV spectra, it is also becoming possible to measure iron abundances in high-$z$ star-forming galaxies \citep[e.g.,][]{steidel2016,cullen2019,topping2020,cullen2021}. These studies represent a promising direction for future high-$z$ galaxy research, because understanding the abundance \emph{patterns} of galaxies provides additional insight regarding their assembly histories---particularly the comparison between iron and $\alpha$ elements (including Mg and O), as the $\alpha$/Fe ratio is sensitive to star formation timescales \citep[e.g.,][]{tinsley1979}.

The goal of this paper is to investigate the abundance patterns of a large sample of individual high-$z$ galaxies in a way that avoids the systematic biases introduced by using locally-calibrated strong-line methods. We update the method introduced by \citet[][hereafter Strom18]{strom2018} and use it to self-consistently infer O/H, N/H, and Fe/H for 195 star-forming galaxies at $z\simeq2-2.7$, the largest sample of high-$z$ galaxies for which multiple elemental abundances have been reported. We introduce the galaxy sample and observations in Section~\ref{sec:data}. Section~\ref{sec:galdna} presents the photoionization model method used for measuring chemical abundances. In Section~\ref{sec:omzr}, we report the M$_{\ast}$-O/H relation for our $z\sim2$ sample and compare and contrast the correlation based on our photoionization model method with the relations from more commonly-used strong-line methods. Section~\ref{sec:newmzr} presents the corresponding M$_{\ast}$-N/H and M$_{\ast}$-Fe/H relations. Section~\ref{sec:patterns} discusses the physical insights that result from measuring abundance patterns rather than a single bulk metallicity, including constraints on galaxy star formation histories. We summarize our findings and conclude in Section~\ref{sec:summary}.

Throughout the paper, we refer to specific spectral features using their vacuum wavelengths and adopt the solar metallicity scale from \citet{asplund2009}, with 
\begin{gather}
12+\log(\textrm{O/H})_{\odot}=8.69 \nonumber \\ 12+\log(\textrm{N/H})_{\odot}=7.83 \nonumber \\
12+\log(\textrm{Fe/H})_{\odot}=7.50 \nonumber
\end{gather}
Historically, 12 is added when reporting log abundances relative to H to ensure that values are always positive. This is not used when reporting abundance ratios such as $\log(\textrm{N/O})$.

\section{Data}
\label{sec:data}

In this work, we analyze a sub-sample of galaxies drawn from the Keck Baryonic Structure Survey \citep[KBSS;][]{rudie2012,steidel2014}, using the same criteria as \citetalias{strom2018} (summarized in Section~\ref{sec:spectroscopy}) to construct the sample. KBSS is a targeted spectroscopic survey of $1.5\lesssim z\lesssim3.5$ galaxies in 15 separate fields, each centered on a bright quasar. Each survey field is approximately $6\arcmin\times8\arcmin$, resulting in a total survey area of $\approx0.24$~deg$^2$. Extensive multiwavelength imaging and both optical and near-infrared (NIR) spectroscopic observations have been conducted in all fields and are described in detail in other work \citep[e.g.,][]{steidel2003,reddy2012,rudie2012,steidel2014,strom2017}. Below, we review aspects of the survey that are most relevant to the present analysis.

\subsection{Photometry and Parent Sample Selection}
\label{sec:photometry}

The majority of KBSS galaxies are selected on the basis of their rest-UV colors, using observed-optical imaging in the $U_n$, $G$, and $\mathcal{R}$ bands and the color selection criteria introduced by \citet{adelberger2004} and \citet{steidel2004} to identify Lyman Break Galaxy analogues at $z\simeq2-2.7$ (sometimes referred to as ``BX" and ``BM" galaxies). This color selection largely translates to a star formation rate-selected sample and is successful at identifying both young galaxies and older galaxies with significant current star formation or with rising star-formation histories (SFHs). Relying on the shape and brightness of galaxies' rest-UV spectral energy distributions (SEDs) can introduce a bias against very massive galaxies and galaxies with heavily reddened ($E(B-V)_{\rm cont}>0.3$) UV spectra. To mitigate these biases, we incorporate knowledge of the shape of galaxies' rest-optical spectra by relaxing the rest-UV color selection and pairing it with a cut in $\mathcal{R}-K_s$ color, which probes the 4000~\AA\, and Balmer breaks across the redshift range of interest. \citet{strom2017} provides a more thorough description of these extended selection criteria, but in brief: these ``RK" galaxies occupy a region of rest-UV color space where $z\sim2$ galaxies selected using other NIR methods, such as distant red galaxies \citep[DRGs,][]{franx2003} and $BzK$ galaxies \citep{daddi2004}, are found \citep[see also the discussion by][]{reddy2005}.

\subsection{Near-infrared Spectroscopy}
\label{sec:spectroscopy}

Beginning in 2012, NIR spectroscopic observations of KBSS galaxies have been conducted in $J$-, $H$-, and $K$-band using the Multi-Object Spectrometer for InfraRed Exploration \citep[MOSFIRE,][]{mclean2012} installed at the Cassegrain focus of the Keck~I telescope on Mauna Kea. We presented the first results from the MOSFIRE component of KBSS in \citet[][hereafter Steidel14]{steidel2014}, with a comprehensive analysis of an expanded sample reported in \citet{strom2017}. We refer the reader to those two papers for complete details regarding the NIR spectroscopic data acquisition, reduction and analysis.

The spectra were acquired using 0\farcs7 slits, with typical seeing of 0\farcs5$-$0\farcs8. The data were reduced using the publicly-available data reduction pipeline\footnote{\url{https://www2.keck.hawaii.edu/inst/mosfire/drp.html}} and subsequently corrected to vacuum heliocentric velocity. If multiple spectra of the same object were acquired, they were shifted to match in the spatial direction and combined using inverse-variance weighting. Slit losses were determined to first-order by comparing observations of a relatively bright star placed on a mask slit with its photometric magnitude in the same band; individual object slit losses were determined by comparing observations of the same object on multiple masks. Typical slit loss corrections are factors of $\sim2$, consistent with estimates based on comparisons of detected continuum with broadband photometry and SED models.

The reduced 2D spectrograms were analyzed using the custom IDL package \texttt{mospec}\footnote{\url{https://github.com/allisonstrom/mospec}}, developed by A. Strom specifically for interacting with MOSFIRE spectra of emission line galaxies \citep{strom2017}. Galaxy spectra were extracted using boxcar extraction apertures, then fit using the best-fit SED model as the stellar continuum and an Gaussian emission line model with a single redshift and line width per band. The ratio of the [\ion{O}{3}]$\lambda\lambda4959,5008$ and [\ion{N}{2}]$\lambda\lambda6549,6585$ doublets are fixed at 3:1, where the longer-wavelength lines are 3$\times$ the strength of their shorter-wavelength counterparts. This greatly improves emission line measurements when one of the two features is impacted by an OH sky line.

To date, $\sim1500$ KBSS galaxies have been observed with MOSFIRE, $\sim800$ of which fall in the redshift range $1.9\leq z\leq 2.7$, where many of the key rest-optical emission features are accessible from the ground at NIR wavelengths. As in \citetalias{strom2018}, galaxies are selected for analysis when there are observations of the spectral regions near H$\alpha$, H$\beta$, [\ion{O}{3}]$\lambda5008$, and [\ion{N}{2}]$\lambda6585$. Measurements of or limits on [\ion{O}{2}]$\lambda\lambda3727,3729$, [\ion{O}{3}]$\lambda4363$, [\ion{S}{2}]$\lambda\lambda6718,6732$, and [\ion{Ne}{3}]$\lambda3869$ are incorporated when present. Objects are included in the sample regardless of the signal-to-noise (S/N) ratio of any single line measurement, but we require the ratio of the band-to-band slit corrections to be less than a factor of 2, as well as a S/N$~>5$ measurement of the Balmer decrement H$\alpha$/H$\beta$, which is used to account for reddening due to dust.\footnote{The S/N calculation for the Balmer decrement accounts for uncertainties in the relative flux calibration between NIR bands.} Galaxies with evidence of significant AGN contamination (in either their rest-UV or rest-optical spectra) or whose H$\alpha$ or H$\beta$ is compromised by a nearby OH line are excluded. In total, these criteria result in a sample of 196 galaxies with $\langle z \rangle = 2.3$. Of these, 195 galaxies have good abundance measurements, as described below in Section~\ref{sec:galdna}.\footnote{It was not possible to identify a photoionization model solution for Q1623-BX449, although the MCMC chains did converge.}

\subsection{Stellar Masses}
\label{sec:masses}

\begin{figure}
\centering
\includegraphics[width=\columnwidth]{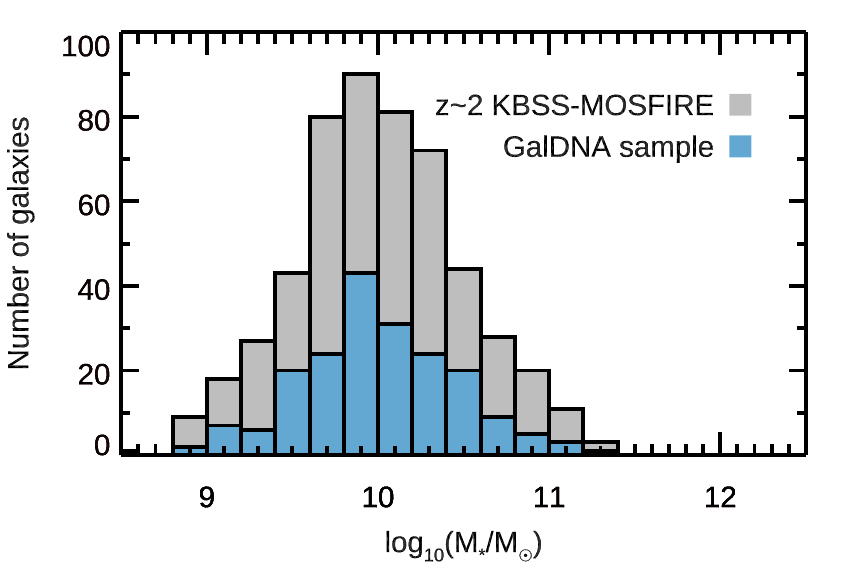}
\caption{The distribution of stellar mass (M$_{\ast}$) for the 195 galaxies in our final sample, shown in blue. For comparison, the grey histogram shows the distribution for the complete parent sample of KBSS galaxies with $1.9\leq z\leq2.7$. The paper sample is consistent with the full KBSS sample and has a median M$_{\ast}=10^{10.0}$~M$_{\odot}$.}
\label{fig:masshist}
\end{figure}

Stellar mass (M$_{\ast}$) estimates for the KBSS galaxies are inferred from reddened stellar population synthesis models fit to broad- and intermediate-band photometry spanning the rest-UV to the rest-IR. The rest-optical (observed-NIR) magnitudes are corrected for line emission using the measurements from the MOSFIRE spectra described above. The general SED fitting methodology is explained by \citet{reddy2012}, with a description of its application to the current KBSS sample found in \citet{strom2017}. \citet{theios2019} examined the impact of using other stellar population models (i.e., those with lower $Z_{\ast}$, as expected for stellar populations at high redshift) on the parameters inferred from SED-fitting. Fortunately, rank-ordering in M$_{\ast}$ is generally preserved regardless of the specific choice of model. As a result, we use the stellar masses from \citet{strom2017}, based on \citet{bruzual2003} stellar population synthesis models with a \citet{chabrier2003} initial mass function, to facilitate a more straightforward comparison with other analyses that use similar SED-fitting techniques and models. The M$_{\ast}$ distributions for the full KBSS sample (grey histogram) and for the 195 galaxies with well-measured abundances (blue histogram) are shown in Figure~\ref{fig:masshist}. These two samples are consistent with being drawn from the same parent population, based on a two-sample Kolmogorov-Smirnov (KS) test. The typical statistical uncertainty on log(M$_{\ast}$/M$_{\odot}$) is $\approx0.16$~dex \citep{shapley2005,erb2006mass}.

\section{Chemical abundances with \texttt{GalDNA}}
\label{sec:galdna}

There are a number of ways to determine the level of chemical enrichment or metallicity of distant galaxies. Broadly speaking, all methods rely on measurements of emission lines from the photoionized gas in galaxies' star-forming regions and fall into two categories: (1) empirical diagnostics based on ``direct" $T_e$-based measurements of a calibration sample where both bright nebular lines and faint auroral lines can be detected or (2) photoionization model methods where observed emission lines are compared to predictions made by models assuming an input ionizing source and set of physical conditions in the gas. 

Empirical methods implicitly depend on the correlation among various quantities including, e.g., the ionization state of the gas and the relative abundances of elements like oxygen and nitrogen. The existence of these underlying correlations causes galaxies and \ion{H}{2} regions to form tight sequences in line ratio space---for example, the ``BPT" diagram comparing [\ion{O}{3}]/H$\beta$ and [\ion{N}{2}]/H$\alpha$ \citep{baldwin1981,veilleux1987}---and allow many different line ratios to be used to infer O/H. Even if a line ratio is only weakly correlated with O/H, so long as the quantity that it \emph{is} most strongly correlated with also correlates with O/H, the diagnostic can still be used. 

In contrast, photoionization model methods allow chosen parameters to vary alongside gas-phase metallicity and thus be explicitly determined for individual objects. Examples include \texttt{IZI} \citep{blanc2015}, which returns ionization parameter $U$ and O/H; \texttt{BOND} \citep{valeasari2016}, which measures both O/H and N/O; and NebulaBayes \citep{thomas2018}, which measures ISM pressure (related to gas density) in addition to $U$ and O/H. Because we are interested in determining multiple chemical abundances for the galaxies in our sample, we use the photoionization model method introduced by \citetalias{strom2018}, which measures O/H, N/H, and Fe/H in addition to $U$. The general method, including recent updates and a revised parameter estimation technique, is described here.

\subsection{Photoionization Model Grid}
\label{sec:model_grid}

We use the same photoionization model predictions for the strong nebular lines in galaxies' rest-optical spectra as \citetalias{strom2018}, which are generated using Cloudy \citep[v13.02;][]{ferland2013}, using stellar population synthesis models from the Binary Population and Spectral Synthesis\footnote{https://bpass.auckland.ac.nz/} code \citep[BPASSv2;][]{eldridge2016,stanway2016} as the input ionizing radiation field. The BPASS models were chosen because of their relative success in reconciling the rest-UV and rest-optical observations of the same $z\sim2$ galaxies, specifically because the implementation of binary evolution physics in BPASS produces harder ionizing radiation fields at fixed $Z_{\ast}$ than, e.g., including stellar rotation or adopting a more top-heavy stellar initial mass function \citep{steidel2016}. This ``boost" of higher energy photons is needed to match the collisional-to-recombination line ratios such as R23\footnote{$\textrm{R23}=\log\left(\frac{[\textrm{\ion{O}{2}}]+[\textrm{\ion{O}{3}}]}{\textrm{H}\beta}\right)$}. In high-$z$ galaxies, these lines ratios tend to be larger than can be explained using stellar models with softer ionizing radiation fields, even at moderate gas-phase O/H corresponding to peak emission from O ions.

We adopt a plane parallel geometry and assume constant star formation histories with an age of 100~Myr and a constant gas density of $n_{\rm H}=300$~cm$^{-3}$. This density was originally chosen to be consistent with the electron density $n_e$\footnote{For ionized gas, $n_{\rm H}\approx n_e$.} determined using the [\ion{O}{2}] ratio measured from a stack of $z\sim2-3$ KBSS galaxies \citep{strom2017}, but is also representative of the typical electron density for \emph{individual} galaxies in the current sample, which ranges from $n_e\simeq150-400$~cm$^{-3}$. Using models with $n_{\rm H}=100$~cm$^{-3}$ or $n_{\rm H}=1000$~cm$^{-3}$ for galaxies with lower or higher densities than this, respectively, does not significantly change their derived abundances or the overall distribution of abundances in the sample. Dust grains are implemented using the ``Orion" mixture, with a dust-to-gas ratio that scales linearly with the metallicity of the gas, $Z_{\rm neb}$. Model grids are calculated for multiple stellar population synthesis models with increasing stellar enrichment ($Z_{\ast}$), allowing $\log(Z_{\rm neb}/Z_{\odot})$ and $\log(U)$, defined as $\log(n_{\gamma}/n_\textrm{H})$, to vary in 0.1~dex steps. Initially, we assume a solar abundance pattern in the gas, but interpolate between the fixed grid points and scale the strength of the nitrogen lines \textit{a posteriori} to accommodate non-solar N/O. These choices allow us to predict emission line intensities (relative to H$\beta$) for the following parameter space:
\begin{eqnarray}
&Z_{\ast}=[0.001,0.014] \textrm{ or } Z_{\ast}/Z_{\odot}\approx[0.07,1.00], \nonumber \\
&Z_{\rm neb}/Z_{\odot}=[0.1,2.0], \nonumber \\
&\log(U)=[-3.5,-1.5], \nonumber \\
&\log({\rm N/O})\geq-2.0. \nonumber
\end{eqnarray}
The lower limit on log(N/O) is $\sim0.5$~dex lower than the primary plateau observed for local \ion{H}{2} regions \citep[e.g.,][]{vanzee1998,izotov1999}.

\begin{figure*}
\centering
\includegraphics[width=\columnwidth]{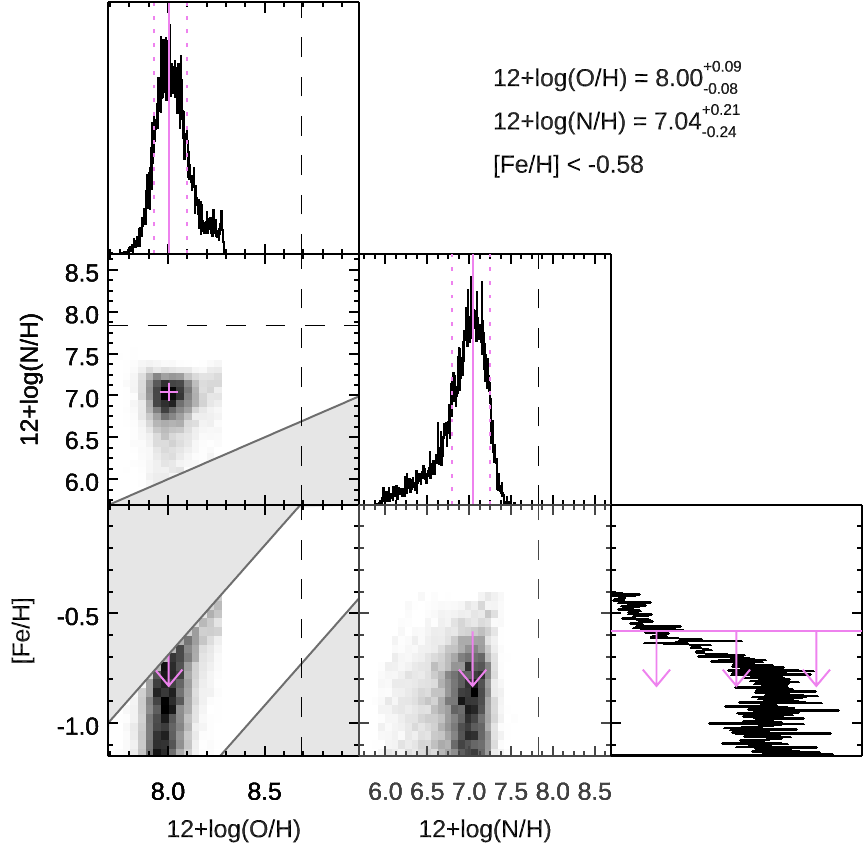}
\includegraphics[width=\columnwidth]{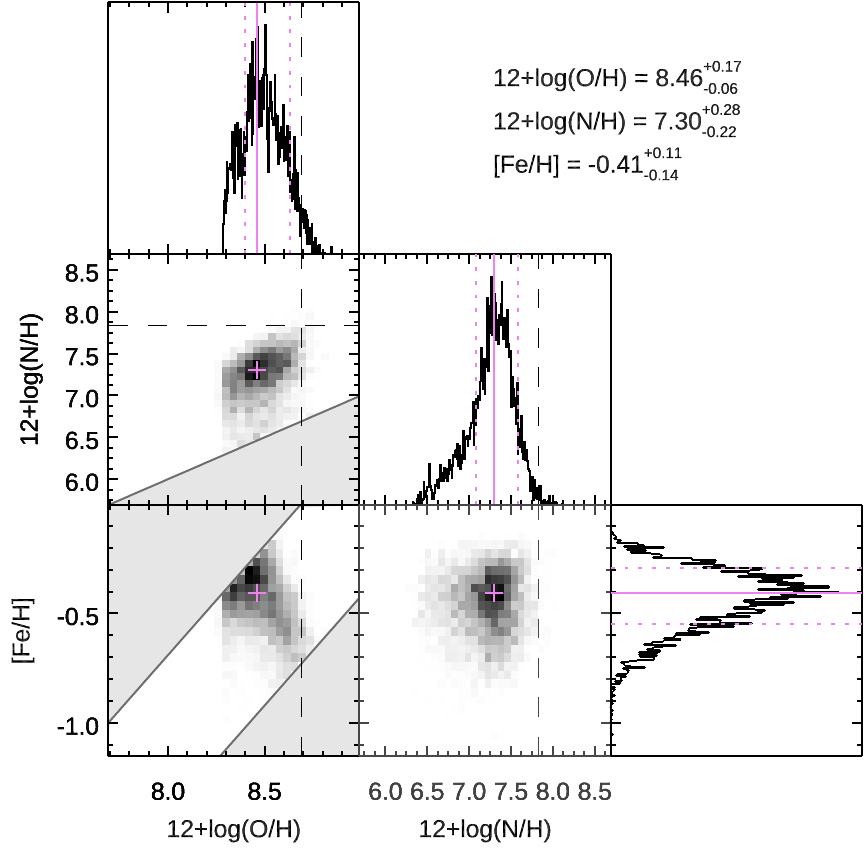}
\caption{The 1D and 2D posterior probability distributions for 12+log(O/H), 12+log(N/H), and [Fe/H] for Q0105-BX147 (M$_{\ast}=10^{9.5}$~M$_{\odot}$), where the low-O/H and high-O/H peaks have been separated into the left and right panels, respectively. These portions of the posterior are considered separately when estimating the abundance pattern in each case (denoted by the purple symbols and solid lines, with the dashed lines indicating the 68\% HDIs). Disallowed regions in N/O and O/Fe are illustrated by the solid grey shading (c.f. Section~\ref{sec:model_grid}), and the solar values of O/H and N/H are identified by the dashed black lines. To decide between potential solutions, we utilize the corresponding ionization parameter estimates to place the galaxy close to the $z\sim2$ locus in mass-ionization space (Figure~\ref{fig:logu_vs_mass}). In this case, the low-O/H solution is preferred.}
\label{fig:two_peaks}
\end{figure*}

These base model parameters can then be converted to abundances:
\begin{eqnarray}
 [\rm Fe/H] &=& \log(Z_{\ast}/Z_{\odot}) \nonumber \\
12+\log(\rm O/H) &=& \log(Z_{\rm neb}/Z_{\odot})+8.69 \nonumber \\
12+\log(\rm N/H) &=& \log(Z_{\rm neb}/Z_{\odot})+8.69+\log(\rm N/O), \nonumber
\end{eqnarray}
where the bracket notation for Fe refers to the enrichment relative to solar, $[\rm Fe/H]=\log(\rm Fe/H)-\log(\rm Fe/H)_{\odot}$. These translations are motivated by several considerations: first, the effect of changing $Z_{\ast}$ on the emission line predictions results from differences in the shape of the ionizing radiation field for a given stellar model family, and the shape of the ionizing spectrum of massive stars is most impacted by the amount of Fe present in their atmospheres. Because this method relies on the Fe enrichment in the massive stars, we avoid concerns about depletion onto dust that impact the abundance of Fe in the gas. In contrast to the important role Fe plays in determining the properties of the massive stars and the amount of gas \emph{heating}, it is O that provides the majority of the gas \emph{cooling} via collisionally-excited lines, due to its relatively high abundance with respect to other elements and many low-lying transitions. Thus, we can assume that changes in $Z_{\ast}$ correspond to changes in Fe/H, and changes in $Z_{\rm neb}$ correspond to changes in O/H. 

We use a Markov chain Monte Carlo (MCMC) approach to sample the posterior within multivariate parameter space (c.f. Section~3.2 of \citetalias{strom2018}), adopting the additional prior that $0.0\leq \log(Z_{\textrm {neb}}/Z_{\ast}) \leq 0.73$. These limits roughly correspond to a range of $\alpha$-enhancement (i.e., O/Fe) between the solar value and Fe-poor CCSNe yields \citep{nomoto2006}.

The photoionization model grids and IDL routines used to implement this method, which we collectively refer to as \texttt{GalDNA} (referencing the analogy of abundance patterns as ``galaxy DNA") are publicly available\footnote{\url{https://github.com/allisonstrom/galdna}}. We caution, however, that some of the built-in assumptions, including that all of the emission from ionized gas observed in a galaxy's spectrum is produced in \ion{H}{2} regions, may be inappropriate for some samples. This is especially true for galaxies at lower redshifts and those with lower SFR surface densities, where contributions from, e.g., diffuse ionized gas and shocks are likely to be non-negligible \citep{sanders2017}.

\subsection{Parameter Estimation}
\label{sec:param_est}

In \citetalias{strom2018}, we reported the maximum \textit{a posteriori} (MAP) values for the four univariate posteriors and required the associated 68\% highest density intervals (HDIs) to be separated from the boundaries of allowed parameter space for all parameters except N/O. Here, we revisit the manner in which upper limits on both N/H and Fe/H are determined, as well as how we treat galaxies with multiple peaks in their posterior posterior probability distribution function (PDF). These changes allow us to better capture the abundance patterns in the largest possible sample.

\subsubsection{Bimodal posteriors}

The connection between specific emission lines and the model parameters in \texttt{GalDNA} is discussed at length in \citetalias{strom2018}, but there are a few key results that inform our physical intuition about how differences in galaxy spectra correspond to differences in the model parameters and, thus, differences in chemical abundances. Foremost is the fact that ratios of collisionally-excited lines to recombination lines (e.g., [\ion{O}{3}]/H$\beta$ and [\ion{S}{2}]$\lambda\lambda6718,6732$/H$\alpha$) effectively trace the amount of photoionization heating per ionization of an H atom. Thus, all else being equal, these ratios will \emph{increase} with \emph{decreasing} $Z_{\ast}$ (i.e., Fe/H), because Fe-poor stars produce more energetic ionizing photons on average than Fe-rich stars. At the same time, many of these ratios are also double-valued: they increase as gas enrichment ($Z_{\rm neb}$ or, equivalently, O/H) increases, until gas-cooling mediated by the heavy elements in the gas cause the gas temperature to decrease, at which point these line ratios also decrease\footnote{A notable exception is [\ion{N}{2}]$\lambda6585$/H$\alpha$, which continues to increase with increasing O/H due to the rapid increase in N/O at higher O/H.}. This behavior is well-known and is one of the challenges inherent in using line diagnostics like R23 to infer O/H, because a single value of R23 can correspond to two different values of O/H. Further, R23 remains at $\sim0.8-1.0$ (which is where most high-$z$ galaxies fall) for a large range in O/H near the ``turnover" where gas cooling begins to outpace increasing gas enrichment. The combination of these effects mean that maximal values of line ratios like [\ion{O}{3}]$\lambda5008$/H$\beta$, [\ion{O}{2}]/H$\beta$, and [\ion{S}{2}]/H$\beta$ are only achieved at low Fe/H \emph{and} moderate O/H. Lower values of these line ratios will usually correspond to either lower or higher O/H, which manifest as bimodal posteriors in \texttt{GalDNA}. Figure~\ref{fig:two_peaks} the posterior PDFs for one galaxy (Q0105-BX147) where this occurs, with the total posterior divided into low-O/H (left panel) and high-O/H (right panel) ``solutions."

\begin{figure}
\includegraphics[width=\columnwidth]{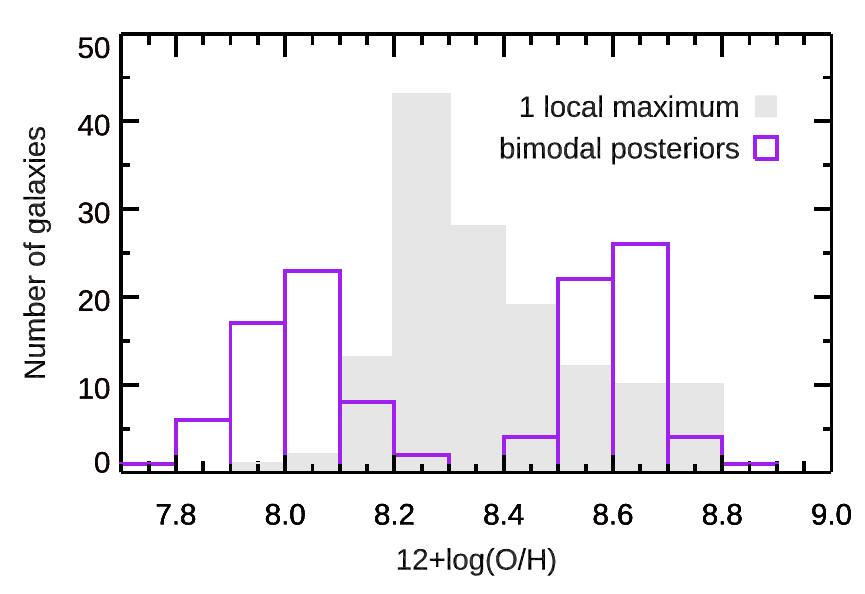}
\caption{The distribution of \texttt{GalDNA}-inferred O/H for the $z\sim2$ galaxies in our sample. The light grey histogram represents the set of 138 objects having either a single local maximum in the posterior PDF or only one valid solution, if the posterior PDF is double-peaked. The open purple histogram shows the distribution of model-inferred O/H for \emph{both} solutions in the 57 remaining galaxies with bimodal posteriors. While the distribution for galaxies with single-peaked posteriors extends to relatively high O/H, the lowest abundances are only probed by galaxies with more than one potential model solution.}
\label{fig:oh_dist}
\end{figure}

Figure~\ref{fig:oh_dist} demonstrates why properly accounting for galaxies with bimodal posteriors is important for characterizing the correlation between O/H and other quantities, such as M$_{\ast}$ or strong-line ratios. Galaxies with a single local maximum in their O/H posterior (the grey histogram) range from $12+\log(\textrm{O/H})\approx8$ (roughly 20\% solar) to just above solar O/H. There is overlap between the upper edge of this distribution and the higher-O/H maxima for galaxies with bimodal posteriors (the open purple histogram). In contrast, lower-O/H maxima only occur in bimodal posteriors and are not seen for galaxies with single-peaked posteriors. 

\begin{figure}
\includegraphics[width=\columnwidth]{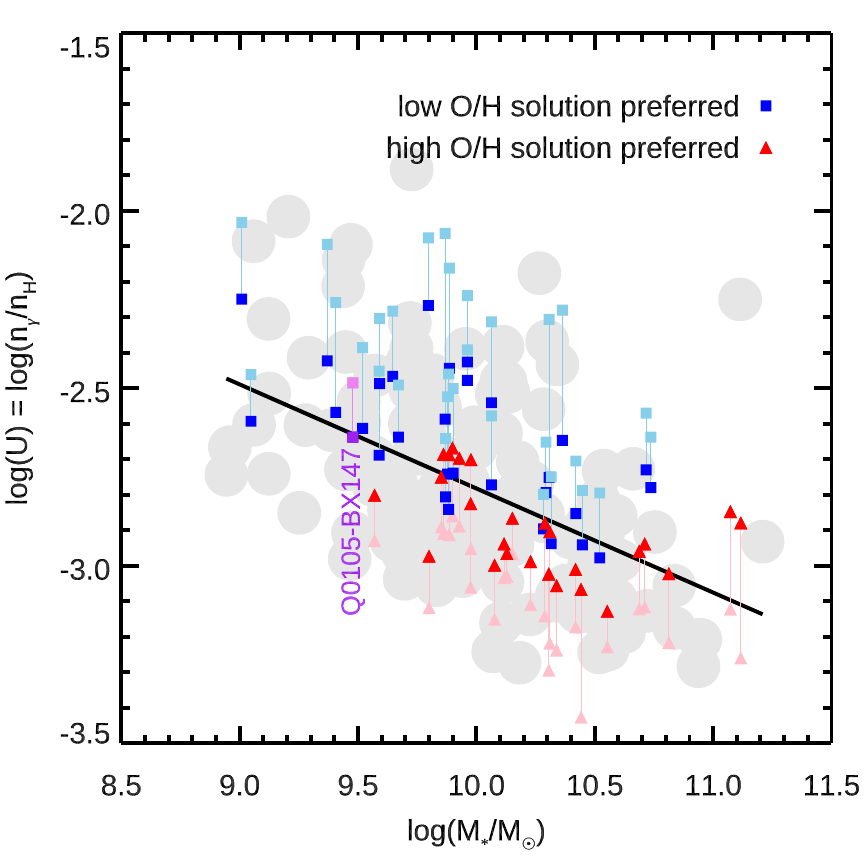}
\caption{Model-inferred $\log(U)$ and M$_{\ast}$ for galaxies with a single peak in the posterior PDF (large grey points) and for galaxies with multi-modal posteriors (colored points). The distributions of both quantities are consistent between the two samples. The best-fit linear relation for the locus of grey points is shown by the solid black line. For galaxies with multiple possible abundance patterns, the darker symbols indicate the solution that is most consistent with this locus, and the lighter symbol shows the location of the second, dispreferred solution. In general, a lower $U$ corresponds to the low-O/H peak in the posterior, and a higher $U$ corresponds to the high-O/H peak in the posterior. The parameter values for Q0105-BX147, which were also shown in Figure~\ref{fig:two_peaks}, are marked by the purple squares; lower $U$ is preferred and corresponds to the lower O/H peak in the posterior (left panel in Figure~\ref{fig:two_peaks}).}
\label{fig:logu_vs_mass}
\end{figure}

To determine which combination of parameters---and thus which O/H---is preferred for individual galaxies, we rely on the anti-correlation between $U$ and M$_{\ast}$. This relation is one of the strongest and most significant correlations among galaxy properties and has been observed in galaxies samples at multiple redshifts using both ionization-sensitive line ratios \citep{sanders2015,sanders2020sulfur,jeong2020} and inferred ionization parameter \citep[e.g.,][]{kaasinen2018}. The tight locus of galaxies in mass-excitation diagram \citep[or MEx; e.g.,][]{juneau2011,juneau2014}, which compares [\ion{O}{3}]/H$\beta$ and M$_{\ast}$, is another manifestation of this relationship.

Using ionization parameter is also practical because it avoids directly imposing a prior on the distribution of abundances and because $U$ is generally the most precisely-determined model parameter when using \texttt{GalDNA}. Further, the distributions of M$_{\ast}$ and $U$ for galaxies with a single maximum in their posterior and for galaxies with bimodal posteriors (including both possible combinations of parameters) are statistically consistent with one another, unlike the distributions of O/H seen in Figure~\ref{fig:oh_dist}. In other words, not only can both sets of galaxies be expected to follow the same locus in the M$_{\ast}$-$U$ plane, but the two $U$ values for galaxies with bimodal posteriors are distinct enough to meaningfully discriminate between the possible combinations of parameters. To accomplish this, we use the single-peaked galaxies to define the best-fit M$_{\ast}$-$U$ relation, shown by the solid black line in Figure~\ref{fig:logu_vs_mass}. For galaxies with bimodal posteriors, we then choose the combination of parameters with $U$ closest to this locus. The preferred ``solutions" determined in this manner are represented by the dark blue squares (low-O/H maxima) and red triangles (high-O/H maxima).

The resulting distributions of $U$ for the low-O/H and high-O/H solutions are significantly different from one another, with the low-O/H galaxies having generally higher $U$ than high-O/H galaxies. For galaxies that are outliers with respect to these $U$ clusters (Q0821-MD5, Q1700-MD103, and Q1623-BX293), we choose the combination of parameters that minimizes this separation. There are 57 galaxies with bimodal posteriors; we select the low-O/H solution for 31 of these and the high-O/H solution for 26. The portion of the multivariate posterior corresponding to the preferred peak is analyzed on its own to determine the MAP values for O/H, N/H, and Fe/H for individual galaxies.

We also investigated other criteria for deciding between unique combinations of parameters for galaxies with bimodal posteriors. These strategies, including assessing the proximity to the N/H-O/H locus of local \ion{H}{2} regions or deciding based on posterior density alone, do not change the qualitative results presented below. We adopt and endorse the method based on the M$_{\ast}$-$U$ relation because it is easily applied in all cases and uses additional knowledge about the individual galaxies themselves to make a more informed choice.

\subsubsection{Nitrogen and Iron Limits}
\label{sec:limits}

Occasionally, the [\ion{N}{2}]$\lambda\lambda6549,6585$ doublet is poorly detected or undetected in the spectra of galaxies in our sample, even when other emission lines (including nearby H$\alpha$) are well-detected. This can occur as a result of low nitrogen abundance and/or high levels of ionization. For our analysis, we do not require the significant detection of any single emission line (just a S/N~$\geq5$ measurement of the Balmer decrement), but these low S/N measurements of [\ion{N}{2}] correspond to posterior PDFs that abut the lower limit of $\log(\textrm{N/O})\geq-2$. Because the imposed prior is on N/O and not N/H, the lowest allowed N/H is a function of O/H. This will tend to produce a well-defined local maximum in N/H space, even when N/O is an upper limit. To account for this, we test whether \emph{both} the N/H and N/O posterior PDFs are bounded (i.e., detached from the limits of the sampled parameter space). If neither the N/H nor the N/O posteriors are bound, or if the N/O posterior alone is not bound, the nitrogen abundance is assumed to be an upper limit given by
\begin{equation}
12+\log(\textrm{N/H}) = 12+\log(\textrm{O/H})+\log(\textrm{N/O})_{68}, \nonumber
\end{equation}
where $\log(\textrm{N/O})_{68}$ is the 68th percentile in log(N/O). Of the 195 galaxies in our sample, 36 have upper limits on N/H determined in this way.

Iron abundance can also be more difficult to estimate because of the more complicated nature of the Fe/H and Fe/H-O/H posteriors. The most common appearance is a horseshoe shape in Fe/H vs. O/H space, opening downward in the lower left panels of the corner plots in Figure~\ref{fig:two_peaks}.  This structure in the 2D posterior arises because there is an inherent trade-off between O and Fe for a given galaxy spectrum: as the ionizing radation field softens with increasing Fe/H---moving toward lower O/Fe---it is necessary to have more O to produce the same line emission but eventually gas cooling due to the increasing O/H begins to dominate, and harder ionizing radiation is required counteract this effect---moving back toward higher O/Fe. This is closely related to the double-valued behavior of R23 with O/H, but additionally highlights the crucial interplay between gas heating \emph{and} gas cooling.

Because of the prior requiring $0.0\leq\textrm{[O/Fe]}\leq0.73$, the two arms of the horseshoe are commonly separated and appear as a bimodal O/H posterior, as we have previously discussed. The 1D Fe/H posterior is also frequently bimodal, although the local maxima are less likely to be bounded. An example of this can be seen in the low-O/H panel on the left in Figure~\ref{fig:two_peaks}, where the marginalized PDF in the bottom right histogram has high posterior density even at the lowest allowed Fe/H, and the 2D posteriors involving Fe/H appear as vertical stripes. In these cases, which also occur at higher O/H, we adopt the 68th percentile in [Fe/H] as an upper limit. A smaller number of galaxies have posteriors that abut the low O/Fe boundary at [O/Fe]$=0.0$, but these often have clear local maxima in the univariate Fe/H posterior, and so we treat them as bounded. In total, 24 galaxies have upper limits on Fe/H.

\section{\texorpdfstring{The $\langle z \rangle = 2.3$ Stellar Mass-O/H Relation}{The <z>=2.3 Stellar Mass-O/H Relation}}
\label{sec:omzr}

Most studies of metallicity in star-forming galaxies, which are largely studied via the nebular emission from their \ion{H}{2} regions, focus on oxygen. This is both physically-motivated---oxygen is the most abundant element (by mass) in the universe after hydrogen and helium---and practical, as emission lines of various ions of oxygen are among the brightest features observed in the spectra of galaxies at all redshifts. As a result, many studies use ``metallicity" interchangeably with oxygen abundance, and the relation between galaxies' stellar mass and gas-phase O/H is among the most common scaling relations discussed in the literature.

\begin{figure}
\centering
\includegraphics[width=\columnwidth]{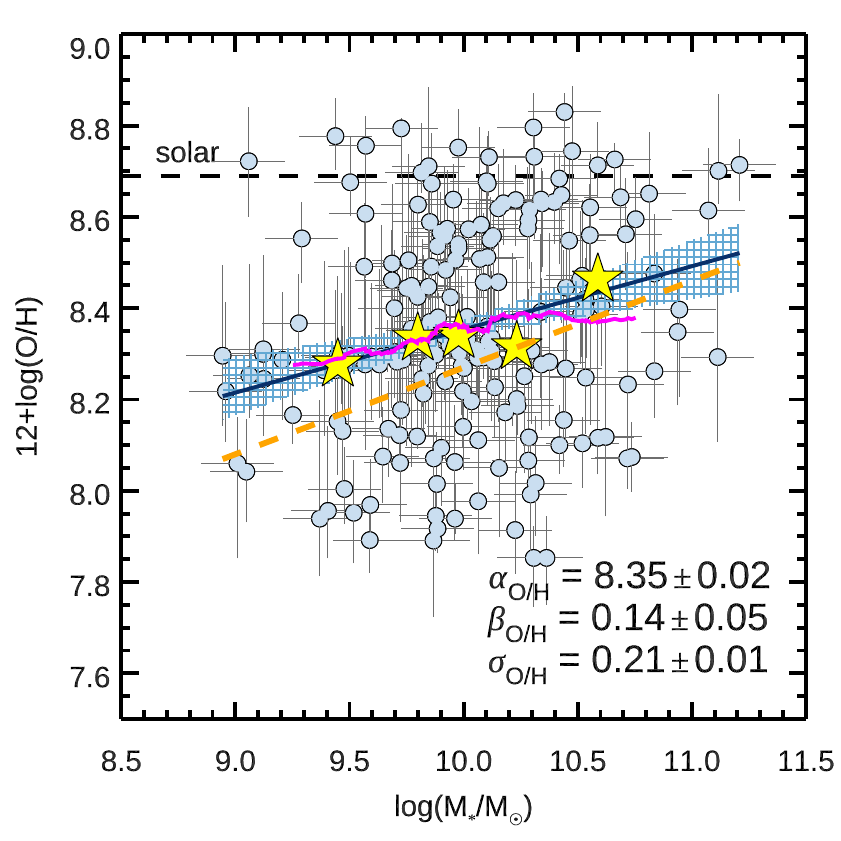}
\caption{The correlation between stellar mass M$_{\ast}$ and \texttt{GalDNA}-based gas-phase O/H for 195 galaxies at $z\sim2$, with the medians in bins of M$_{\ast}$ represented by the yellow stars. The solar value of $12+\log(\textrm{O/H})_{\odot}=8.69$ is represented by the horizontal dashed line. The magenta line shows the center moving average for O/H with M$_{\ast}$. The dark blue line is the best-fit linear relation, with the blue hatching indicating the 1$\sigma$ uncertainties on the slope and normalization. Most $z\sim2$ galaxies have sub-solar O/H, and there is a marginally-significant positive correlation between O/H and M$_{\ast}$ (Spearman $\rho=0.21$, $p=0.003$). However, there is also substantial intrinsic scatter ($\sigma_{\textrm{O/H}}=0.21\pm0.01$~dex), which together with the measurement uncertainties could camouflage a more significant correlation. The O3N2-based O-MZR reported by \citetalias{steidel2014} for a smaller sample of KBSS galaxies is shown by the dashed orange line.}
\label{fig:omzr}
\end{figure}

\begin{deluxetable*}{@{\extracolsep{5pt}}lrcrrrr}
\tablewidth{0pt}
\tablecaption{Abundance scaling relations with M$_{\ast}$ \label{tab:params}}
\tablehead{
\colhead{} & \multicolumn{2}{c}{Spearman test} & \multicolumn{4}{c}{\texttt{LINMIX\_ERR}} \\
\cline{2-3} \cline{4-7}
\colhead{Element} & \colhead{$\rho$} & \colhead{$p$} & \colhead{P(\%)\tablenotemark{a}} & \colhead{$\alpha$\tablenotemark{b}} & \colhead{$\beta$} & \colhead{$\sigma_\textrm{int}$ (dex)}
}
\startdata
12+log(O/H)$_\textrm{GalDNA}$ & 0.21 & $3\times10^{-3}$ & 99.8 &$8.35\pm0.02$ & $0.14\pm0.05$ & $0.21\pm0.01$ \\
12+log(O/H)$_\textrm{O3N2}$ & 0.50 & $9\times10^{-14}$ & \dots & \dots & \dots & \dots \\
12+log(N/H) & 0.28 & $8\times10^{-5}$ & 100.0 & $7.07\pm0.03$ & $0.29\pm0.07$ & $0.33\pm0.02$ \\
$[\textrm{Fe/H}]$ & 0.27 & $1\times10^{-4}$ & 100.0 & $-0.69\pm0.02$ & $0.17\pm0.05$ & $0.22\pm0.01$ \\
\enddata
\tablenotetext{a}{Probability of a positive linear relation, based on the posterior distribution for $\beta$}
\tablenotetext{b}{Determined at M$_{\ast}=10^{10}$~M$_{\odot}$, c.f. Equation~\ref{eq:mzr}}
\end{deluxetable*}

Figure~\ref{fig:omzr} shows the observed correlation between O/H and M$_{\ast}$ (hereafter O-MZR) for our sample of 195 $z\sim2$ galaxies, where O/H has been measured using \texttt{GalDNA} as described in Section~\ref{sec:galdna}. The asymmetric vertical error bars for each point represent the 68\% highest density interval (HDI) on the univariate O/H posterior, and the horizontal error bars represent the typical $\approx0.16$~dex uncertainty on the stellar masses, not taking into account any systematic uncertainties in determining M$_{\ast}$. A Spearman rank correlation test indicates a weak ($\rho=0.21$) but significant ($p=0.003$) positive correlation\footnote{For the purposes of characterizing correlations in this paper, we consider $\rho<0.3$ to be a weak correlation, $0.3\leq\rho\leq0.6$ to be a moderate correlation, and $\rho>0.6$ to be a strong correlation. We use the propability of the null result $p$ to assess the significance of the correlation, with $p\leq0.003$ considered significant.} (Table~\ref{tab:params}). The magenta line shows the center moving average to give a non-parametric sense of this correlation, which increases by $\approx0.12$~dex in $\log(\textrm{O/H})$ (or $\sim30\%$ in O/H) over 1.5 decades in stellar mass. The yellow stars represent the median $12+\log(\textrm{O/H})$ in bins of M$_{\ast}$, which also show an increase in oxygen enrichment with increasing mass.

The large apparent scatter (due to both measurement uncertainties and intrinsic scatter) could camouflage a more significant correlation than indicated by the Spearman test. To account for this, we also use \texttt{LINMIX\_ERR} \citep{kelly2007} to characterize the relation between M$_{\ast}$ and O/H; \texttt{LINMIX\_ERR} is an IDL-based routine that performs linear regression when there are measurement errors in both variables. Using this method, we find a $99.8\%$ probability of a positive correlation, compared to $99.7\%$ from the Spearman test. The solid blue line shows the best-fit linear relation,
\begin{equation}
12+\log(\textrm{O/H}) = \alpha+\beta[\log(\textrm{M}_{\ast}/\textrm{M}_{\odot})-10],
\label{eq:mzr}
\end{equation}
with $\alpha_\textrm{O/H}=8.35\pm0.02$ and $\beta_\textrm{O/H}=0.14\pm0.05$. These fit parameters, as well as the intrinsic scatter of this relation ($\sigma_{\textrm{O/H}}=0.21\pm0.01$), are also reported in Table~\ref{tab:params}. The hatched blue area represents the $1\sigma$ uncertainties in the slope and normalization.

The general sense of this relation is consistent with other studies of the O-MZR at $z\sim2-3$. The majority of galaxies have sub-solar oxygen abundance, with $\approx45\%$ (O/H)$_{\odot}$ at M$_{\ast} = 10^{10}$~M$_{\odot}$ and trending toward solar at high M$_{\ast}$. For comparison, the first O-MZR reported for KBSS galaxies by \citetalias{steidel2014}, which used the O3N2\footnote{$\textrm{O3N2}=\log([\textrm{\ion{O}{3}}]/\textrm{H}\beta)-\log([\textrm{\ion{N}{2}}]/\textrm{H}\alpha)$ } diagnostic from \citet[][hereafter PP04]{pettini2004} to infer O/H, is shown by the dashed orange line in Figure~\ref{fig:omzr} and is defined by $\alpha_\textrm{Steidel14}=8.27\pm0.01$ and $\beta_\textrm{Steidel14}=0.19\pm0.02$. The new \texttt{GalDNA} relation is shallower than the \citetalias{steidel2014} relation, with the slopes differing by a little more than $1\sigma$. The normalization of the \texttt{GalDNA} O-MZR is also higher than the earlier O3N2 relation by $0.08\pm0.02$~dex. These differences can be attributed to changes in the sample and sample size, as well as differences in methodology, which we discuss in more detail below.

In particular, discrepancies in the abundance scale or normalization of the O-MZR determined using different methods are not unexpected. The O3N2 calibration from \citetalias{pettini2004} that is used in \citetalias{steidel2014} is based on a sample of nearby \ion{H}{2} regions with known $T_e$-based abundances. However, the direct method has been shown to result in abundances up to $\simeq0.24$~dex \emph{lower} than those measured using recombination lines \citep{esteban2004,esteban2014,blanc2015}. \citet{steidel2016} found an offset of $0.25$~dex between the $T_e$-based O/H and the photoionization model-inferred O/H for a composite spectrum of 30 KBSS star-forming galaxies; the O/H estimated using the O3N2 index was intermediate between these two values. The offset between the new \texttt{GalDNA} O-MZR and the older O3N2 O-MZR from \citetalias{steidel2014} is less than the offset between the model O/H and O3N2 O/H reported in that work.

\subsection{Slope}
\label{sec:omzr_slope}

The \texttt{GalDNA} O-MZR in Figure~\ref{fig:omzr} has $\beta_{\textrm{O/H}}=0.14\pm0.05$ and is shallower than the previously-reported KBSS O-MZR from \citetalias{steidel2014}, as well as most other high-$z$ O-MZR relations reported in the literature \citep[][among others]{erb2006metal,maiolino2008,yabe2014,zahid2014,sanders2018,sanders2021}, which range from $\beta\sim0.2$ to $\beta\sim0.35$. Our current understanding of galaxy assembly reasonably predicts a positive correlation between M$_{\ast}$ and O/H, given that the more stars that are created, the larger the mass of oxygen is produced. The slope of the O-MZR can also be affected by other astrophysical processes, including inflows and outflows, which determine how much of the oxygen is retained or expelled and how much it is diluted by infalling low-metallicity gas. A comparatively shallow O-MZR may reflect a relationship between gas content and M$_{\ast}$ at $z\sim2$ that keeps O/H higher in low-M$_{\ast}$ galaxies and/or leads to relatively lower O/H in high-M$_{\ast}$ galaxies. For example, a shallower O-MZR at high redshift could result if low-M$_{\ast}$ galaxies are not efficiently ejecting their enriched gas or if high-M$_{\ast}$ galaxies retain more of their original (pristine) gas mass or accrete more low-metallicity gas from their surroundings, both of which would dilute the oxygen. The effects of these processes are explored by multiple cosmological simulations \citep[e.g.,][]{brooks2007,ma2016mzr,derossi2017,torrey2019}, and we briefly return to this issue in Section~\ref{sec:patterns}, where we discuss the O-MZR together with the scaling relations for nitrogen and iron.

However, before considering the astrophysical interpretation of the O-MZR, we must first consider the practical issue of converting observed quantities to O/H and how that may affect the resulting relation. For example, a shallow O-MZR is a natural consequence of a relatively small dynamic range observed in line ratios like R23 in $z\sim2$ samples. However, other line ratios (including O3N2) exhibit a stronger correlation with M$_{\ast}$, even in high-$z$ samples, and these have historically been used to measure the O-MZR instead. As we have discussed, these O-MZR relations can be dramatically different from one another, but these differences should not always be interpreted as having \emph{physical} meaning, because one of the primary drivers of the slope of strong-line relations is the calibration used to infer O/H.

\begin{figure}
\centering
\includegraphics[width=\columnwidth]{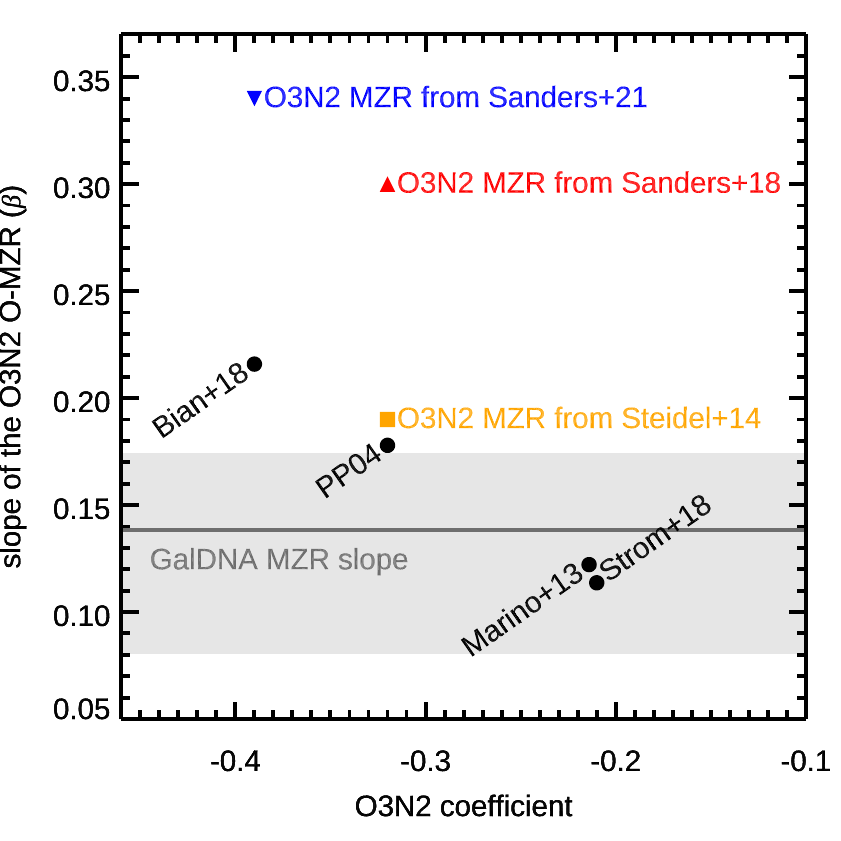}
\caption{The slope of the O3N2 O-MZR is extremely sensitive to the choice of O3N2 calibration, which is in turn dependent on the correlations among parameters in the original calibration sample. The round points represent the O3N2 O-MZR measured for our $z\sim2$ galaxy sample using four different calibrations. As the slope of the O3N2 calibration becomes more negative, the slope of the O-MZR grows steeper, which can result in a steeper O-MZR than observed in Figure~\ref{fig:omzr} (horizontal grey line and shaded region). Three O3N2 O-MZRs from the literature are shown by the orange square \citep[][based on a smaller sample of KBSS galaxies]{steidel2014}, the red triangle \citep[from MOSDEF;][]{sanders2018}, and the inverted blue triangle \citep[][also from MOSDEF]{sanders2021}. These points illustrate the differences that can arise merely from sample size and sample selection. Notably, the magnitude of the differences that result from choosing different diagnostics is similar to or even larger than sample effects.}
\label{fig:omzr_slope}
\end{figure}

Figure~\ref{fig:omzr_slope} shows how different O3N2 calibrations with different coefficients for the line ratio can result in very different O-MZR slopes. The black points represent the slope of the O3N2 O-MZR measured when different calibrations for O3N2 (identified by the labels) are used to infer O/H for the same sample of galaxies as shown in Figures~\ref{fig:masshist} and \ref{fig:omzr}; these results are also summarized in 
Table~\ref{tab:slopes}. Comparing the two most dissimilar calibrations, from \citetalias{strom2018} and \citet{bian2018}, shows that the slope of the O-MZR can differ by up to $\Delta_{\beta}\approx0.1$ \emph{only because a different calibration was used}. One may attempt to circumvent this uncertainty by selecting the calibration deemed most appropriate for the sample being studied, but it is not always clear how that decision should be made. \citet{bian2018} use direct-method abundances from stacks of ``local analogues" to high-$z$ galaxies to construct their calibration. \citet{marino2013} also use the direct method, in combination with a multiple-line-ratio technique, for \ion{H}{2} complexes in local galaxies that were observed as part of the CALIFA survey \citep{sanchez2012}. In \citetalias{strom2018}, we constructed an O3N2 calibration based on the photoionization model oxygen abundances for our $z\sim2$ sample (albeit only the sub-sample with single peaks), which resulted in an equation with a coefficient very similar to one based on \ion{H}{2} regions from \citet{pilyugin2012}. In the absence of a sufficient sample of direct-method abundances in \emph{high-redshift} galaxies, any or all of these calibrations might be good choices, depending on the sample under consideration.

\begin{deluxetable}{lcrl}
\tablewidth{0pt}
\tablecaption{O3N2 O-MZR slopes \label{tab:slopes}}
\tablehead{\colhead{Sample} & \colhead{O3N2 calibration} & \colhead{Coefficient} & \colhead{Slope}}
\startdata
this paper & PP04 & $-0.32$ & $0.18\pm0.02$ \\
this paper & Marino+13 & $-0.214$ & $0.12\pm0.01$ \\
this paper & Bian+18 & $-0.39$ & $0.22\pm0.02$ \\
this paper & Strom+18 & $-0.21$ & $0.11\pm0.01$ \\
\hline
Steidel+14 & PP04 & $-0.32$ & $0.19\pm0.02$ \\
Sanders+18 & PP04 & $-0.32$ & $0.30$ \\
Sanders+21 & Bian+18 & $-0.39$ & $0.34$ \\
\enddata
\end{deluxetable}

The difference between the \texttt{GalDNA} O-MZR and the \citetalias{pettini2004} O3N2 O-MZR for our current sample is $\Delta_{\beta}=0.04$. This is larger than the difference in slope when the same \citetalias{pettini2004} O3N2 calibration is used for both the current sample and the sample from \citetalias{steidel2014} ($\Delta_{\beta}=0.01$). We therefore conclude that, in this case, the principal explanation for the difference in slope between our current analysis and \citetalias{steidel2014} seen in Figure~\ref{fig:omzr} is the choice of abundance inference technique, with sample effects playing a more minor role.

At the same time, it is worth noting that the slope of the O3N2 O-MZR reported for KBSS by \citetalias{steidel2014} (the orange square in Figure~\ref{fig:omzr_slope}) and the slope reported for MOSDEF by \citet[][the red triangle in Figure~\ref{fig:omzr_slope}]{sanders2018} is still quite large ($\Delta_{\beta}=0.11$), even though both are based on the \citetalias{pettini2004} calibration for O3N2. This indicates that differences in sample \emph{selection} rather than sample \emph{size} can also play a role. Nonetheless, because it is impossible to know \textit{a priori} the specific origin(s) of any reported differences in the O-MZR, it is likely more useful to compare the scaling relations between M$_{\ast}$ and \emph{line ratio}---rather than between M$_{\ast}$ and O/H, if O/H has been inferred using a strong-line diagnostic.

\subsection{Intrinsic Scatter}
\label{sec:omzr_intsc}

\begin{figure*}
\centering
\includegraphics{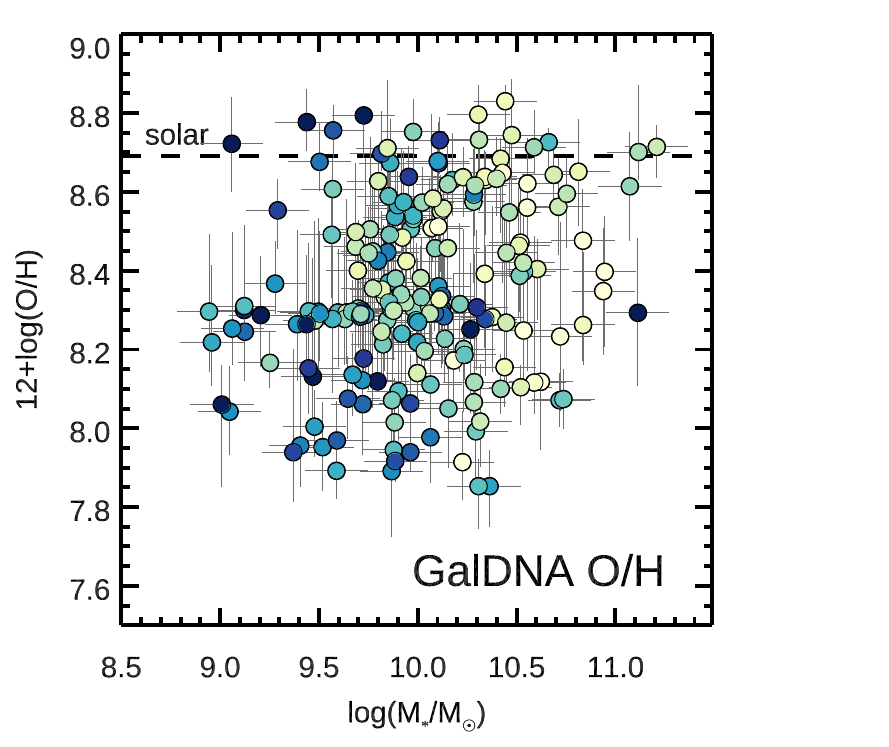}
\hspace{-1.28in}
\includegraphics{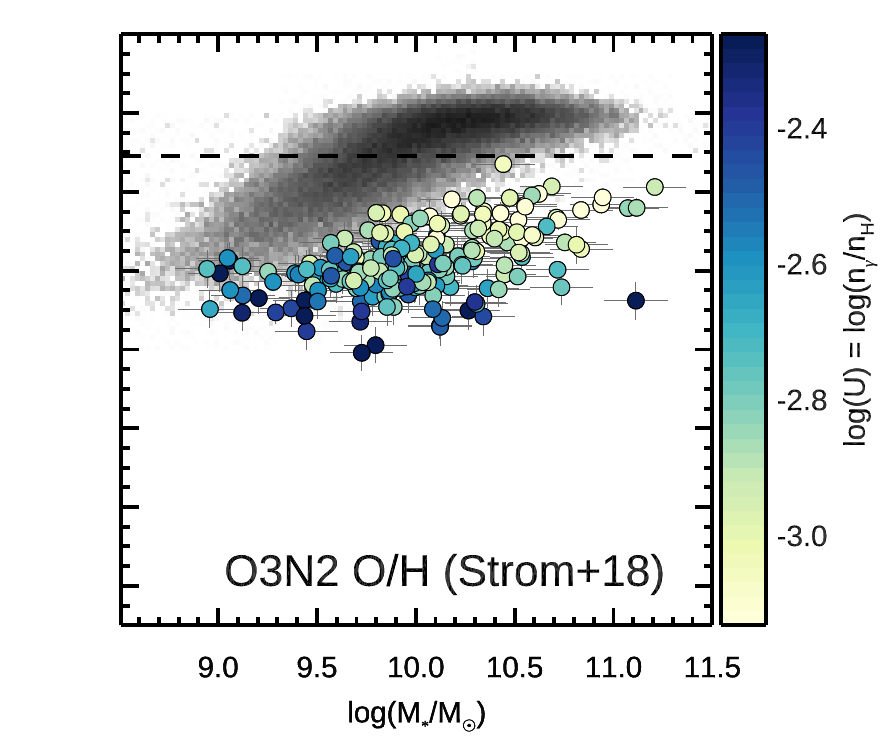}
\caption{The O-MZR for the $z\sim2$ galaxy sample presented in this paper, color-coded by ionization parameter (yellow corresponds to low $U$, and blue corresponds to high $U$). The oxygen abundances were determined using \texttt{GalDNA} in the left panel and using the O3N2 diagnostic from \citetalias{strom2018} in the right panel (the distribution of $12+\log(\textrm{O/H})_{\textrm{O3N2}}$ for $z\sim0$ SDSS galaxies is shown in greyscale). The correlation between O/H and M$_{\ast}$ is statistically significant in both cases, but a difference in intrinsic scatter leads to remarkably different appearance in the two panels. A strong correspondence between O3N2 and $U$, which manifests as horizontal isocontours in the right panel, is fundamental to the construction of strong-line diagnostics at $z\sim0$. Consequently, the tighter O3N2 O-MZR is the result of effectively measuring a combination of two parameters---$U$ \emph{and} O/H---and not due to low intrinsic scatter in the O-MZR itself. This trend is not observed in the \texttt{GalDNA} O-MZR in the left panel, as O/H and $U$ are only weakly anti-correlated using the model method.}
\label{fig:omzr_intsc}
\end{figure*}

A notable feature of the \texttt{GalDNA} O-MZR presented in Figure~\ref{fig:omzr} is the substantial scatter, with the intrinsic scatter (after accounting for measurement error) estimated to be $\sigma_{\textrm{O/H}}=0.21\pm0.05$~dex. This is reasonably consistent with the $0.15$~dex intrinsic scatter reported for the $z\sim0$ O-MZR for galaxies with similar M$_{\ast}$ \citep{tremonti2004} but is more visually apparent here because of the comparatively larger O/H measurement uncertainties for the high-$z$ galaxies.

The origin of the scatter in the O-MZR is still a matter of some debate, although there is a general consensus that deviations from the $z\sim0$ relation correlate with SFR, with higher-SFR galaxies exhibiting lower O/H at fixed M$_{\ast}$ \citep[e.g.,][]{ellison2008,lara-lopez2010}. This ``fundamental metallicity relation" \citep[FMR;][]{mannucci2010} is likely the result of differences in gas fraction \citep{bothwell2013}, in the sense that higher-SFR galaxies have larger gas fractions and, thus, have diluted the O formed by their stellar populations. The scatter may also be attributed to the characteristic timescale for oxygen enrichment \citep[][see also Section~\ref{sec:patterns}]{matthee2018}, gas flows \citep{vanloon2021}, or observational effects like inclination angle \citep{tremonti2004}.

Accurately characterizing the intrinsic scatter in chemical scaling relations like the O-MZR is critical, as it may provide insight regarding the importance of these other processes and/or effects. However, as with the slope of such relations, measuring intrinsic scatter is very sensitive to the method used to infer abundances \citep{tremonti2004}. For example, the intrinsic scatter we report for the \texttt{GalDNA} O-MZR is significantly larger than reported for the O-MZR when strong-line diagnostics are used to infer O/H, which is usually $\sigma_{\textrm{O/H}}\lesssim0.08$~dex. However, given the similar (or even larger) scatter in the measurements originally used to establish the diagnostics, such a small amount of scatter is unexpected---and may instead indicate that the O-MZR determined using strong-lines is a projection of a higher-dimensional surface linking M$_{\ast}$, O/H, and other parameters.

Figure~\ref{fig:omzr_intsc} shows that ionization parameter $U$ is a likely candidate for a third parameter that would account for the diminution in observed scatter. The left panel shows the \texttt{GalDNA} O-MZR from Figure~\ref{fig:omzr}, and the right panel shows the O3N2 O-MZR for the same galaxies using the O3N2 calibration from \citetalias{strom2018}. In both panels, points have been color-coded based on the value of $U$ for that galaxy. The trend with $U$ is remarkably different between the panels: the isocontours of $U$ are nearly horizontal in the right panel, while there is considerable scatter in $U$ in the left panel. Using the Pearson correlation coefficients, we can compare the strength of the correlation between model parameters from \texttt{GalDNA} and O3N2: we find $r_{\textrm{O/H-O3N2}}=-0.40$, indicating a moderate anti-correlation between O3N2 and log(O/H), whereas there is a much stronger correlation between O3N2 and log($U$), $r_{U\textrm{-O3N2}}=0.74$. The correlation between O3N2 and \emph{both} $12+\log(\textrm{O/H})$ and $\log(U)$ is $r_{\textrm{O/H+}U}=0.78$. This reflects the fact that O3N2 calibrations depend on a strong underlying (anti-)correlation between O/H and $U$. We did not find strong evidence for such an anti-correlation in our $z\sim2.3$ sample in \citetalias{strom2018}, but our new results for galaxies with bimodal posteriors indicate that the galaxies with higher $U$ tend to have lower O/H and vice versa, although it is not clear whether this is consistent with $z\sim0$ $U$-O/H relation underpinning all locally-calibrated diagnostics. Even if the $U$-O/H relation is the same at $z\sim2$ as at $z\sim0$, the strong correlation between O3N2 and $U$ in most calibration samples means that the scatter in \emph{any} O3N2 O-MZR is not accurately tracing the scatter in the underlying O/H scaling relation. The additional dependence on $U$ will artificially suppress the apparent scatter and thus the inferred intrinsic scatter as well \citep[c.f. Section~7.2,][]{steidel2014}.

Trends with ionization parameter could be the result of underlying differences in ionizing photon flux density, in \ion{H}{2} region gas density and pressure, or both. However, we find no significant correlation between M$_{\ast}$ and $n_e$ in our sample (Spearman $\rho=-0.06$, $p=0.43$), suggesting that differences in $n_{\gamma}$ may be more important. The role of ionization parameter, its relationship to other quantities such as SFR, and how it should be physically interpreted in high-$z$ galaxies will be addressed in future work. Here, we emphasize that methods like \texttt{GalDNA} that explicitly account for other key parameters---including $U$, but also multiple abundances---offer a promising opportunity to investigate scaling relations in greater detail and build more astrophysical intuition from the same observations.

\subsection{Measuring Shallow Relations with Large Scatter}
\label{sec:shallow}

\begin{figure}
\centering
\includegraphics[width=\columnwidth]{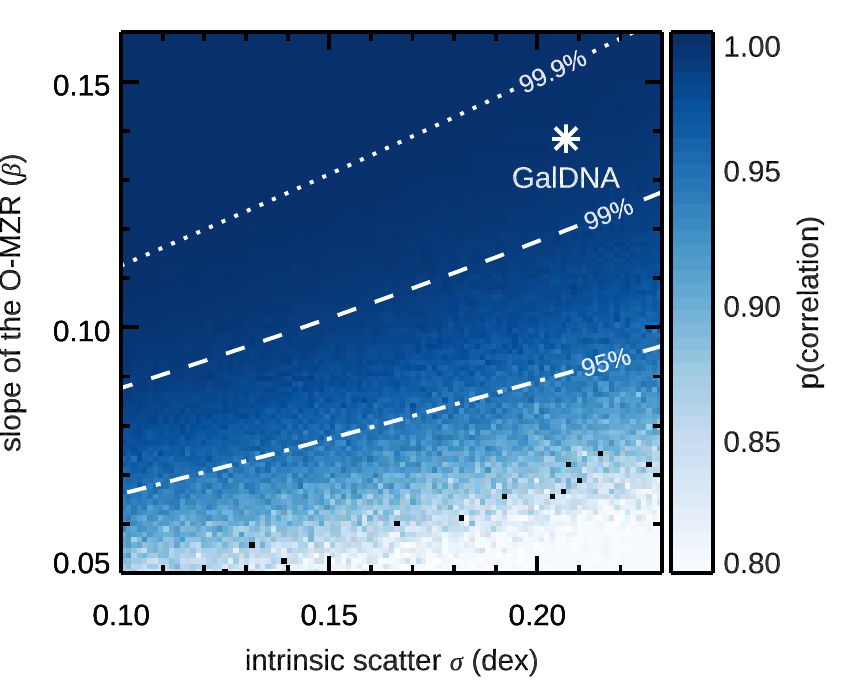}
\includegraphics[width=\columnwidth]{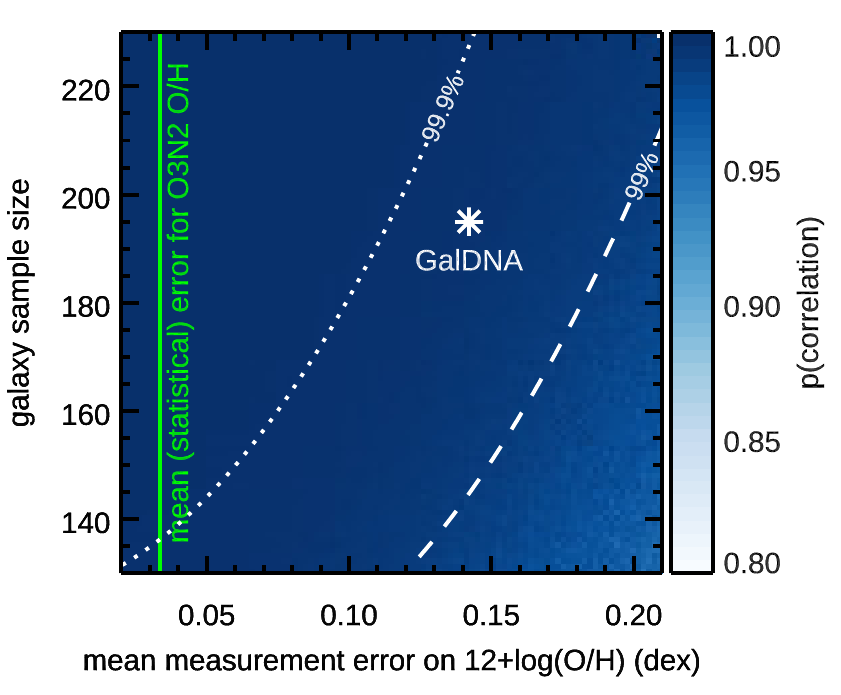}
\caption{The significance of a monotonic correlation between O/H and M$_{\ast}$, assessed using a Spearman rank correlation test, as a function of the intrinsic scatter and slope of the underlying O-MZR (top) and as a function of the mean measurement error and galaxy sample size (bottom). Darker colors indicate a more significant correlation, and the $95\%$, $99\%$, and $99.9\%$ confidence isocontours are denoted by the dashed and dotted white lines. As seen in the top panel, the intrinsic slope of the O-MZR has a stronger effect on the significance of the observed relation than differences in intrinsic scatter (for a fixed sample size and mean measurement error). Assuming the intrinsic O-MZR has the comparatively shallow slope and moderate intrinsic scatter measured from Figure~\ref{fig:omzr}, improving the precision of abundance measurements could dramatically increase the significance of the observed scaling relation. The mean $1\sigma$ error on $12+\log(\textrm{O/H})_\textrm{O3N2}$ (vertical green line, bottom panel) is small enough to ensure a highly significant observed correlation even for much smaller samples---but this does not include systematic uncertainties stemming from scatter in the original calibration and is likely to be an underestimate of the true uncertainty on O3N2-based O/H abundances.}
\label{fig:recover}
\end{figure}

Above, we highlighted scenarios that might result in a shallow O-MZR with relatively large scatter. However, it is much more challenging to confidently characterize such relations than steeper or tighter correlations.

To understand our results in the context of other O-MZR measured for similar samples, we simulated the correlation between O/H and M$_{\ast}$ for a sample of 195 galaxies with the same distribution of M$_{\ast}$ as shown by the blue histogram in Figure~\ref{fig:masshist}. We consider a range of slopes $\beta_{\textrm{O/H}}=0.05-0.16$ and intrinsic scatter $\sigma_{\textrm{O/H}}=0.10-0.23$~dex. The Spearman rank correlation coefficient and the probability of the null result are then calculated for $n=500$ realizations for each $\beta_{\textrm{O/H}}$ and $\sigma_{\textrm{O/H}}$.

The top panel of Figure~\ref{fig:recover} shows the median correlation significance ($1-p_{\textrm{Spearman}}$) for the O-MZR as a function of slope $\beta_{\textrm{O/H}}$ and intrinsic scatter $\sigma_{\textrm{O/H}}$, with darker shading indicating a more significant correlation. The dashed and dotted white lines denote the $95\%$, $99\%$, and $99.9\%$ confidence isocontours. The location of the \texttt{GalDNA} O-MZR in this parameter space is marked by the white star (with $p=0.997$). This experiment demonstrates why the strong-line-determined O-MZRs from the literature and the O3N2 O-MZR in Figure~\ref{fig:omzr_intsc} appear more obvious visually and tend to be more statistically significant. Such relations are generally steeper and have smaller scatter, which would position them to the upper left in comparison to the corresponding confidence isocontours. As we have noted, however, it is not obvious that the observed scatter in the O3N2 O-MZR reflects the true intrinsic scatter.

The two most straightforward ways to increase the significance of an observed correlation with a given intrinsic $\beta$ and $\sigma$ are to increase the sample size or decrease the statistical measurement uncertainties. The bottom panel shows the results from a similar experiment as the previous one, but now varying the galaxy sample size (where the stellar masses are drawn from a normal distribution with the same mean and variance as the paper sample) and mean measurement error on the abundance. For reference, the mean measurement error on $\log(\textrm{O/H})_{\textrm{O3N2}}$ is shown by the vertical green line, although this does not capture any systematic uncertainties. A modest increase in sample size (to $\sim230$~galaxies) or decrease in measurement error (to $\sim0.11$~dex) would increase our confidence in a positive correlation to $99.9\%+$.

In coming years, we will continue to see progress on both fronts: samples of high-quality rest-optical spectra of $z\gtrsim2$ galaxies are increasing in size, and advances in models of massive star evolution and independent constraints on Fe/H in high-$z$ galaxies will both lead to narrower posterior distributions for O/H. In the meantime, it is important to continue to invest resources in developing methods of abundance inference that can accurately characterize both the slope \emph{and} the scatter in the O-MZR. It is also important to study galaxies' abundance \emph{patterns} rather than a single proxy for their bulk metallicity. \texttt{GalDNA} is an example of this kind of method, and in the next section, we present the corresponding scaling relations for nitrogen (N-MZR) and iron (Fe-MZR) that result from our analysis. 

\section{Relations for Other Elements}
\label{sec:newmzr}

Here we present the N-MZR and Fe-MZR for the same galaxies that define the O-MZR in Figure~\ref{fig:omzr}. These results comprise the largest study to date of multiple elemental abundances in individual high-$z$ galaxies.

\subsection{The N-MZR}
\label{sec:nmzr}

\begin{figure}
\centering
\includegraphics[width=\columnwidth]{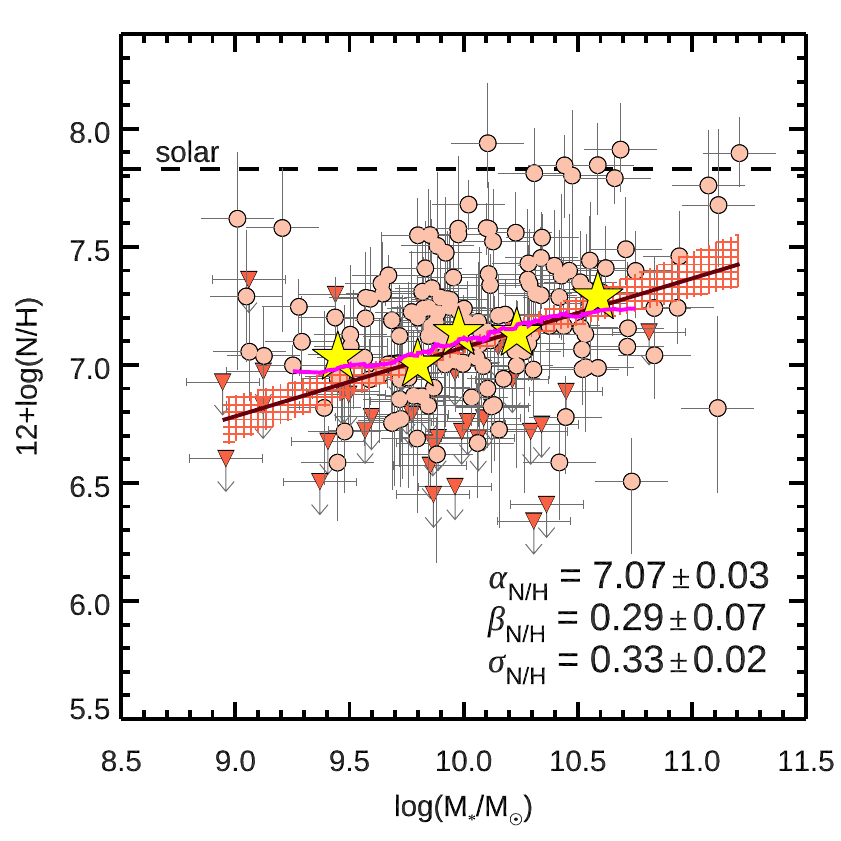}
\caption{The correlation between stellar mass M$_{\ast}$ and gas-phase N/H for our $z\sim2$ galaxy sample, shown in the same manner as Figure~\ref{fig:omzr}. Upper limits on N/H are shown as darker red triangles. The magenta line follows the center moving average, and the solid red line indicates the best-fit linear relation (uncertainties illustrated by the hatched region). Medians in bins of M$_{\ast}$ are shown by the yellow stars. The observed correlation is stronger than the O-MZR, with a rank correlation coefficient of $\rho=0.28$ ($p=8\times10^{-5}$) and a slope of $\beta_{\textrm{N/H}}=0.29\pm0.07$ for the linear relation. The intrinsic scatter in this relation is also larger than that measured for the O-MZR ($\sigma_{\textrm{N/H}}=0.33\pm0.02$~dex).}
\label{fig:nmzr}
\end{figure}

The correlation between M$_{\ast}$ and N/H for our sample of 195 $z\sim2$ galaxies is shown in Figure~\ref{fig:nmzr}. As for the O-MZR in Figure~\ref{fig:omzr}, individual galaxies are represented by the colored points, with the darker triangles indicating upper limits on N/H (c.f. Section~\ref{sec:limits}). The center moving average in magenta provides a non-parametric view of the N-MZR, and the yellow stars represent the median $12+\log(\textrm{N/H})$ in bins of M$_{\ast}$. The Spearman rank correlation coefficient for the N-MZR is $\rho_{\textrm{N/H}}=0.28$ ($p=8\times10^{-5}$), indicating a significant moderate correlation. The correspondence with M$_{\ast}$ is stronger for N/H than for O/H, although it also remains sub-solar across the entire sampled mass range, with $12+\log(\textrm{N/H})=7.07\pm0.03$ at $10^{10}$~M$_{\odot}$, or $\approx17\%$ (N/H)$_{\odot}$.

The best-fit linear relation determined using \texttt{LINMIX\_ERR} is shown by the dark red line, with the hatched region illustrating the 1$\sigma$ uncertainties on the slope and normalization; the fit parameters can be found in Table~\ref{tab:params}. The N-MZR is significantly steeper than the O-MZR with $\beta_{\textrm{N/H}}=0.29\pm0.07$ and has significantly larger intrinsic scatter $\sigma_{\textrm{N/H}}=0.33\pm0.02$. The mean measurement error for $12+\log(\textrm{N/H})$ is also larger, with $s_{\textrm{N/H}}=0.23$, in large part due to the fact that N/H is determined almost entirely by observations of a single feature [\ion{N}{2}]$\lambda\lambda6549,6585$ that tracks the number of nitrogen atoms in the gas. This is in contrast to O/H, which dominates the cooling in \ion{H}{2} regions and thus influences all collisionally-excited transitions. As a result, O/H can be inferred from an ensemble of lines. The larger scatter in the N-MZR may have a physical origin as well, as nitrogen is thought to be produced by intermediate mass stars and returned to the ISM on somewhat longer timescales than oxygen \citep{pettini2002}.

\subsection{The Fe-MZR}
\label{sec:femzr}

\begin{figure}
\centering
\includegraphics[width=\columnwidth]{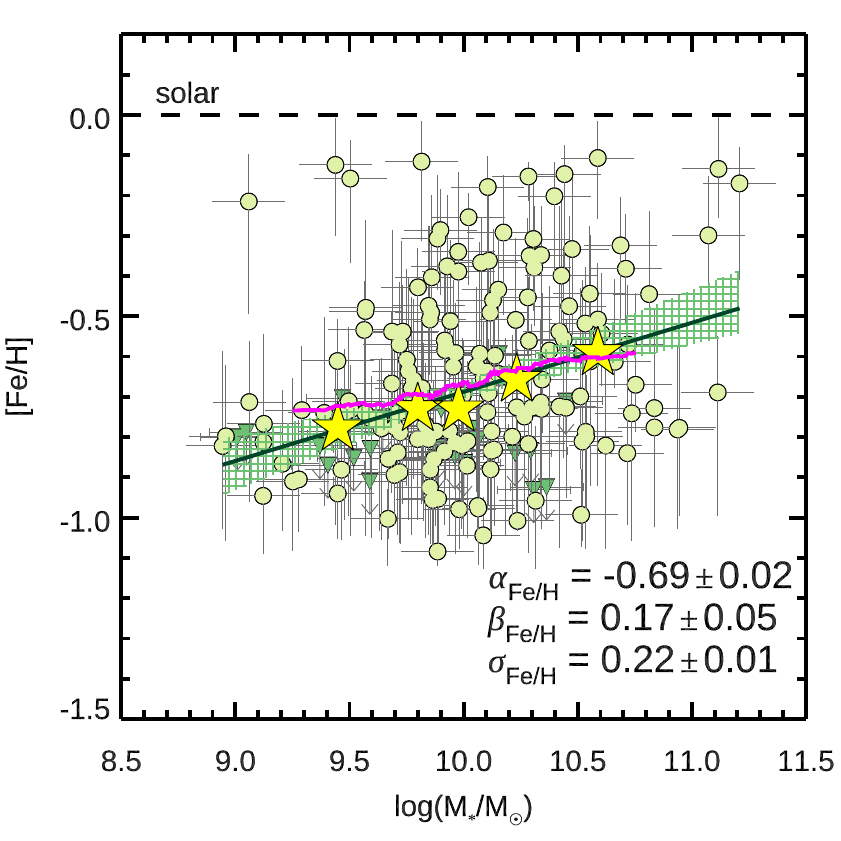}
\caption{The correlation between stellar mass M$_{\ast}$ and Fe/H for our $z\sim2$ galaxy sample. As in Figures~\ref{fig:omzr} and \ref{fig:nmzr}, the magenta line follows the center moving average, the solid green line indicates the best-fit linear relation (uncertainties illustrated by the hatched region), and medians in bins of M$_{\ast}$ are shown by the yellow stars. The rank correlation coefficient is $\rho=0.27$ ($p=1\times10^{-4}$), and the linear relation has $\beta_{\textrm{Fe/H}}=0.17\pm0.05$ and $\sigma_{\textrm{Fe/H}}=0.22\pm0.01$~dex. Unlike the O-MZR and N-MZR, the Fe-MZR remains significantly sub-solar even at high M$_{\ast}$.}
\label{fig:femzr}
\end{figure}

Figure~\ref{fig:femzr} shows the correlation between M$_{\ast}$ and inferred [Fe/H] for our $z\sim2$ galaxy sample in the same manner as the O-MZR in Figure~\ref{fig:omzr} and the N-MZR in Figure~\ref{fig:nmzr}. Rather than present the iron abundances at face value, however, we use bracket notation, which indicates the enrichment relative to solar and is commonly used in studies of stellar enrichment. The Spearman rank correlation coefficient for the Fe-MZR is $\rho_{\textrm{Fe/H}}=0.27$ ($p=1\times10^{-4}$), which indicates a stronger and more significant correlation than the O-MZR for the same sample.

The Fe-MZR also has a comparatively low normalization, with $[\textrm{Fe/H}]=-0.69$ ($\approx20\%$ solar) at M$_{\ast}=10^{10}$~M$_{\odot}$. This is consistent with the stellar metallicity relation (read: Fe-MZR) reported by \citet{cullen2019} for a sample of $2.5\leq z\leq5$ galaxies, which had $[\textrm{Fe/H}]\approx-0.7$ based on measurements from rest-UV spectra. This relatively low degree of enrichment is expected, given that the typical ages of star-forming galaxies at $z\sim2$ ($\lesssim300$~Myr) are lower than the characteristic timescale for substantial iron enrichment due to Type~Ia SNe ($\sim1$~Gyr).

As with the O-MZR, the Fe-MZR in Figure~\ref{fig:femzr} is shallower than other recent studies at $z\gtrsim2$ \citep{cullen2019,cullen2021}, which found $\beta_{\textrm{Fe/H}}\sim0.3$ using galaxies' rest-UV spectra to estimate stellar metallicity. \citet{theios2019} also showed a somewhat steeper increase in stellar metallicity as a function of M$_{\ast}$ ($\beta_{\textrm{Fe/H}}\approx0.29$) for KBSS galaxies based on stacks of rest-UV spectra in bins of stellar mass (see Figure~16 in that paper). Using rest-UV photospheric spectra is a more direct method of inferring stellar metallicity than \texttt{GalDNA}, and we showed in \citet{steidel2016} that stellar metallicity inferred this way is relatively insensitive to the choice of stellar model. However, \citet{cullen2021} found that stellar metallicities inferred using BPASS were systematically lower than metallicities inferred using Starburst99 models \citep{leitherer2010}. There is is also substantial diversity in the predicted ionizing spectra of the same massive stars, which is what \texttt{GalDNA} relies on to indirectly estimate Fe/H. Both methods are ultimately model-dependent---albeit in different ways---and it is currently difficult to directly compare them without larger, representative samples of individual galaxies where stellar metallicity can be measured using rest-UV spectra.

\section{Astrophysical Insight from Abundance Patterns}
\label{sec:patterns}

The principal advantage of using photoionization model-based methods like \texttt{GalDNA} is the ability to explicitly account for additional factors that influence the physics of star-forming regions. Photoionization models are also the best way to constrain abundance ratios for the same galaxies using common observables. Routinely acquiring sufficiently deep rest-UV spectra to measure Fe/H for individual high-$z$ galaxies remains prohibitively expensive, although this will hopefully change in the future with the construction of ``extremely large telescopes" (ELTs).

\begin{figure}
\centering
\includegraphics[width=\columnwidth,trim=0 27 0 55,clip]{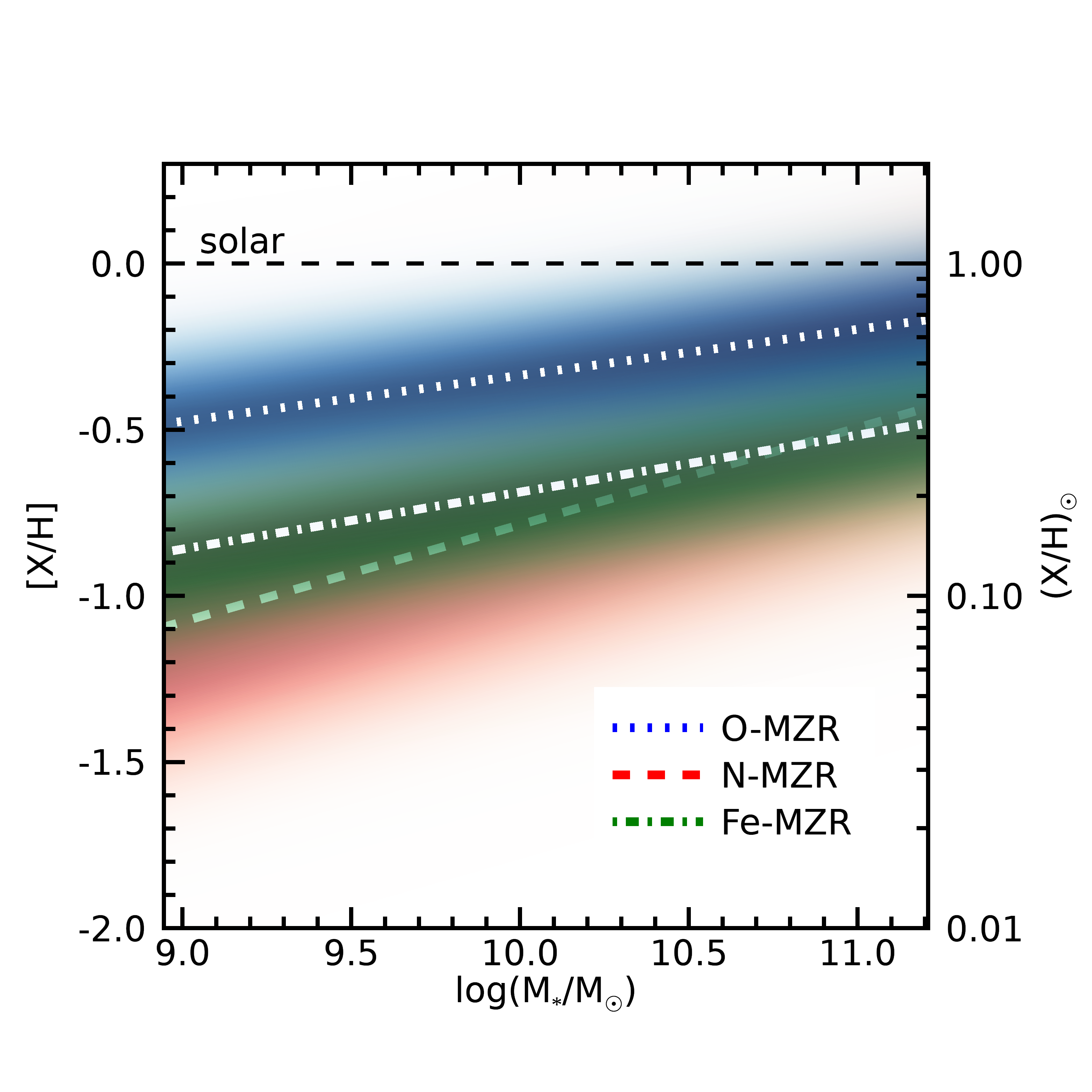}
\caption{A comparison of the O-MZR (blue), N-MZR (red), and Fe-MZR (green) for our $z\sim2$ galaxy sample, plotted on the same log scale relative to solar. The shading illustrates the intrinsic scatter measured for each relation. The O-MZR and Fe-MZR have very similar slopes and are offset from one another by $0.35\pm0.03$~dex at $M_{\ast}=10^{10}$~M$_{\odot}$, which translates to $\textrm{O/Fe}\approx2.2\times(\textrm{O/Fe})_{\odot}$. This stays relatively constant across the sampled mass range, implying that young stars even in massive galaxies at $z\sim2$ are significantly $\alpha$-enhanced. The N-MZR rises much more quickly than the other relations, starting near the primary plateau in N/O at M$_{\ast}\sim10^{9}$~M$_{\odot}$ and surpassing the Fe-MZR near M$_{\ast}\sim10^{11}$~M$_{\odot}$.}
\label{fig:mzrcomp}
\end{figure}

In this section we revisit aspects of the discussion in Sections~\ref{sec:omzr} and \ref{sec:newmzr}, but now consider how the normalization, slope, and scatter of the separate scaling relations can be used \emph{together} to build physical intuition regarding the chemical evolution of high-$z$ galaxies. Figure~\ref{fig:mzrcomp} shows the best-fit O-MZR (blue, dotted white line), N-MZR (red, dashed white line), and Fe-MZR (green, dot-dashed white line) described above on the same log scale relative to solar to facilitate comparison. The shading around the best-fit linear relations indicates the intrinsic scatter in each case.

\subsection{Alpha-enhancement in High-redshift Galaxies}
\label{sec:norm_comp}

\begin{figure}
\centering
\includegraphics[width=\columnwidth]{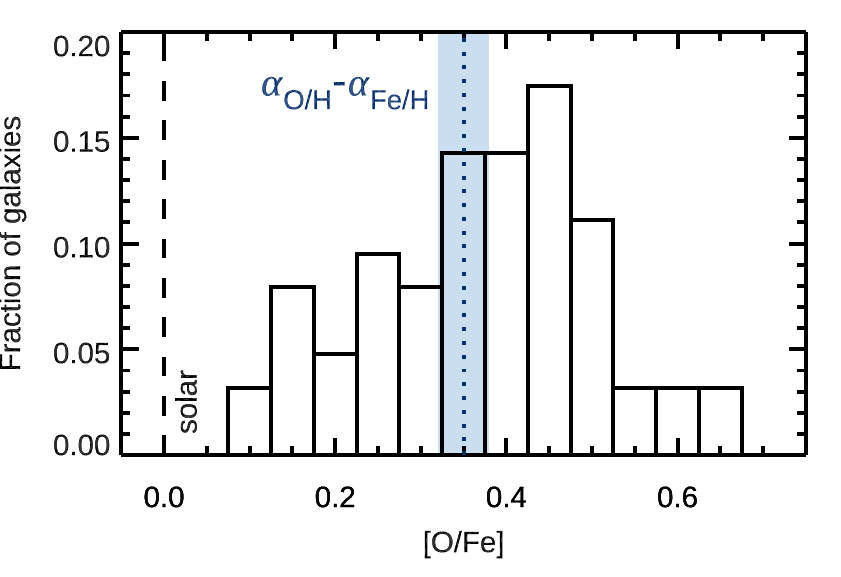}
\caption{The distribution of [O/Fe] for the 63 individual galaxies with bound posteriors in the abundance ratio. For these galaxies, the median [O/Fe]~$=0.39$, corresponding to $\textrm{O/Fe}\approx2.5\times(\textrm{O/Fe})_{\odot}$. This is slightly higher but still consistent with the [O/Fe] inferred from the offset between the O-MZR and Fe-MZR, which is shown by the dotted blue line, with the shaded region representing the statistical uncertainty in this value.}
\label{fig:ofe_hist}
\end{figure}

The offset between the O-MZR and Fe-MZR yields [O/Fe]~$=(\alpha_{\rm O/H}-8.69)-\alpha_{\rm Fe/H}=0.35\pm0.03$ at $M_{\ast}=10^{10}$~M$_{\odot}$, corresponding to $\textrm{O/Fe}\approx2.2\times(\textrm{O/Fe})_{\odot}$. This is lower than the median ratio reported in \citetalias{strom2018}, [O/Fe]~$=0.42$ or $\textrm{O/Fe}\approx2.6\times(\textrm{O/Fe})_{\odot}$, but still consistent given the typical measurement uncertainties ($\sim0.2$~dex). Figure~\ref{fig:ofe_hist} shows the updated distribution of [O/Fe] for the 63 individual galaxies in the current sample with bounded posterior PDFs for O/Fe. These galaxies have a median [O/Fe]~$=0.39$ or $\textrm{O/Fe}\approx2.5\times(\textrm{O/Fe})_{\odot}$ and an interquartile range of [O/Fe]~$=0.28-0.47$ or $\textrm{O/Fe}\approx1.9-3.0\times(\textrm{O/Fe})_{\odot}$. Both estimates of the typical O/Fe in our $z\sim2-3$ galaxy sample are in line with the $\alpha$-enhancement reported by \citet{cullen2019} and \citet{sanders2020}. These works focused on different spectroscopic features from one another and from our analysis, and the agreement between all of these methods should be seen as strong confirmation of elevated $\alpha/$Fe in high-$z$ galaxies.

Although the slope of Fe-MZR is slightly higher than the slope of the O-MZR, they agree within errors, indicating a relatively constant level of $\alpha$-enhancement across the entire sampled mass range. Likewise, there is no statistically significant anti-correlation between [O/Fe] and M$_{\ast}$ for the subset of galaxies in Figure~\ref{fig:ofe_hist} (Spearman $\rho=-0.29$, $p=0.02$). Perhaps the most direct comparison of this finding to other samples is with the recent results for $z\sim3$ galaxies from \citet{cullen2021}, who compared their rest-UV-measured Fe/H with the strong-line-based O-MZR from \citet{sanders2021}. As we do, the authors found virtually identical---albeit much steeper---slopes for their gas-phase metallicity relation (O-MZR) and stellar metallicity relation (Fe-MZR).

The constancy of super-solar O/Fe with M$_{\ast}$ appears robust to the choice of method and is intriguing, as it may suggest that the mode of production for O and Fe remains the same for galaxies spanning the observed range in M$_{\ast}$. The implication then is that the Fe in even relatively massive galaxies has been produced---along with the O---by CCSNe, rather than by Type Ia SNe. Significant Fe production by Type Ia SNe would tend to decrease O/Fe in older galaxies, and we would expect O/Fe to decrease with increasing M$_{\ast}$ (i.e., for the Fe-MZR to be steeper than the O-MZR).

This is contrary to our understanding of $\alpha$-enhancement in nearby galaxies, which is highest in massive galaxies \citep[e.g.,][]{faber1992,thomas2010,conroy2014,segers2016}. However, our analysis pertains to the abundances in and around young, massive stars and captures the \emph{current} level of enrichment in galaxies, whereas the $\alpha$-enhanced stellar populations in, e.g., elliptical galaxies are composed of lower mass stars that likely formed at earlier times, some at similar redshifts as our galaxy sample. Significant $\alpha$-enhancement in $z\sim2$ galaxies can occur if they are relatively young ($\lesssim300$~Myr) or have rising SFHs. Considering the observed trend in cosmic star formation rate density at these redshifts, it is possible that this is the case and that older galaxies and galaxies with declining SFHs---where we might expect lower O/Fe---are comparatively uncommon. However, it may also suggest that these galaxies are underrepresented in existing surveys of emission-line galaxies.

Setting aside differences in the slopes reported for the O-MZR and Fe-MZR in different studies, the slope of the Fe-MZR must effectively function as an upper limit on the slope of the O-MZR. The Fe-MZR may be \emph{steeper}---if Type~Ia SNe have contributed to the iron in the ISM of older, more massive galaxies---but generally not \emph{shallower} than the O-MZR. This is because the limiting case is an enrichment pattern resulting from pure CCSNe ejecta at all stellar masses, which would tie together the slopes of the two relations. Curiously, the typical $\alpha$-enhancement in our sample ([O/Fe]~$=0.35\pm0.03$) is less than half what is expected from pure Fe-poor CCSNe enrichment ([O/Fe]$\approx0.73$; \citealt{nomoto2006}). Some of this discrepancy could result from our definitions of solar enrichment for O and Fe, but it may also point to other sources of enrichment, for example ``prompt" Type Ia SNe with characteristic delay times $\lesssim400$~Myr \citep{scannapieco2005,matteucci2009,yates2013}.

Nitrogen production is clearly not tied to oxygen and iron production in the same way, as the N-MZR has a significantly steeper slope than both the O-MZR and the Fe-MZR (Figure~\ref{fig:mzrcomp}). At the low-mass end of our sample (M$_{\ast}=10^9$~M$_{\odot}$), comparing the O-MZR and N-MZR yields $\log(\textrm{N/O})\sim-1.45$, close to the primary plateau. Based on the slope and normalization of the two relations, $\log(\textrm{N/O})_{\odot}=-0.86$ would occur at M$_{\ast}=10^{12.8}$~M$_{\odot}$, a much higher mass than one might expect to need to reach solar abundances. This may indicate that the \texttt{GalDNA} N-MZR relation in Figure~\ref{fig:nmzr} is actually too shallow.

Still, a lower N/H coupled with a more rapid increase with M$_{\ast}$ relative to O/H is a robust result that persists regardless of how the final sample is constructed, including the treatment of upper limits on N/H and N/O (c.f. Section~\ref{sec:limits}). These relative abundances suggest that most high-$z$ galaxies with M$_{\ast}\gtrsim10^9$~M$_{\odot}$ are engaging in secondary production of nitrogen. In high-$z$ damped Lyman-$\alpha$ systems (DLAs), where nitrogen and oxygen abundances have also been studied in some detail, N/O is near or below the primary plateau at $\log(\textrm{N/O})\sim-1.5$, but these systems are generally very metal-poor and have lower O/H than the galaxies in our sample \citep[e.g.,][]{pettini1995,pettini2002,pettini2008,petitjean2008}. It seems likely that pushing $z\sim2$ galaxy surveys to lower masses would allow us to begin directly studying galaxies producing primary nitrogen.

\subsection{Timescales of Galaxy Assembly}
\label{sec:intsc_comp}

The behavior of the intrinsic scatter of the MZR for different elements can provide meaningful benchmarks for simulations of galaxy formation and evolution. For example, \citet{matthee2018} argued that the scatter in $z=0.1$ chemical scaling relations should be expected to differ depending on the element, with the O-MZR having the least instrinsic scatter and the N-MZR and Fe-MZR having significantly larger scatter. Their interpretation of the physical origin of these differences relates to the different timescales on which elements are formed and returned to the ISM as a result of their progenitors. For $\alpha$ elements like O, which is produced by CCSNe on relatively short timescales, deviations from equilibrium are short-lived and should therefore result in less observed scatter.

\begin{figure}
\centering
\includegraphics[width=\columnwidth]{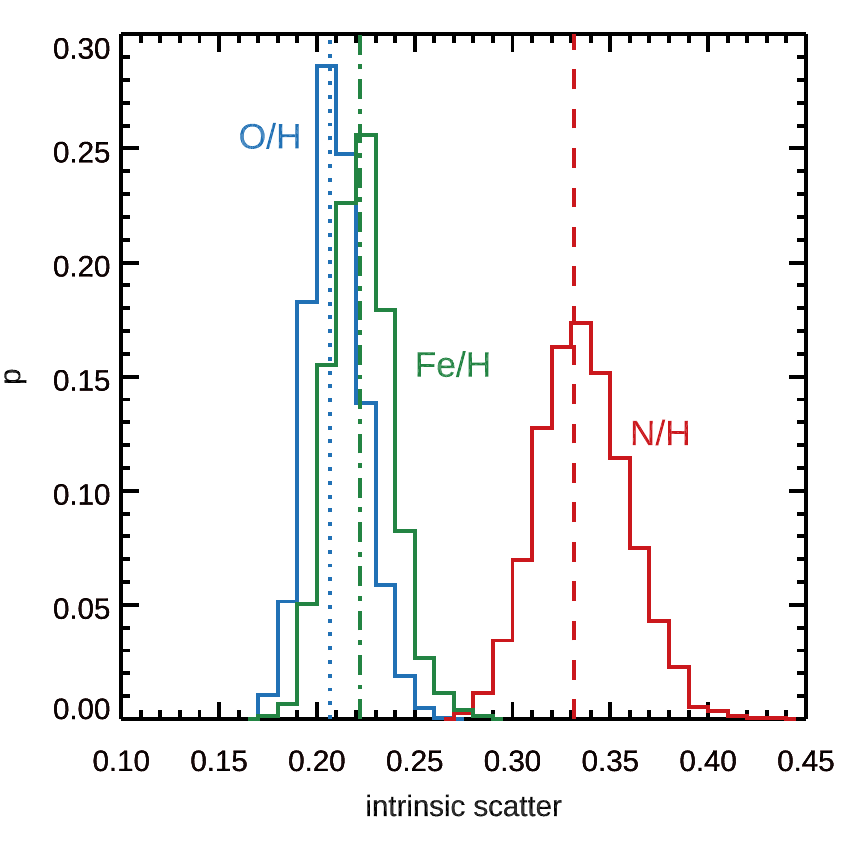}
\caption{The intrinsic scatter in the correlation between M$_{\ast}$ and abundance varies between elements. The histograms show the posterior distributions of $\sigma_{\textrm{int}}$ from fitting the relations plotted in Figures~\ref{fig:omzr}, \ref{fig:nmzr}, and \ref{fig:femzr}. The intrinsic scatter in the O-MZR is the smallest ($\sigma_\textrm{O/H}=0.21\pm0.01$), followed by the Fe-MZR ($\sigma_\textrm{Fe/H}=0.22\pm0.01$), and the N-MZR ($\sigma_\textrm{N/H}=0.33\pm0.02$). The comparatively low scatter in the Fe-MZR relative to the N-MZR and its relative similarity to the O-MZR may be evidence that most or all of the iron in our galaxy sample was created by CCSNe.}
\label{fig:intsc}
\end{figure}

Figure~\ref{fig:intsc} shows the posterior distributions of $\sigma_{\textrm{int}}$ for the O-MZR (blue, dotted line), Fe-MZR (green, dot-dashed line), and N-MZR (red, dashed line). As predicted, the O-MZR exhibits the smallest intrinsic scatter, $\sigma_{\textrm{O/H}}=0.21\pm0.01$. For our sample, however, the intrinsic scatter of the Fe-MZR is $\sigma_{\textrm{Fe/H}}=0.22\pm0.01$~dex, statistically identical to the O-MZR and significantly smaller than that of the N-MZR. The relative scatter suggests that most or all of the iron enrichment in our galaxy sample occurred on comparatively shorter timescales like oxygen (i.e., in CCSNe). Such an interpretation is consistent with the super-solar O/Fe observed for galaxies at all M$_{\ast}$, as discussed previously.

The large intrinsic scatter of the N-MZR $\sigma_{\textrm{N/H}}=0.33\pm0.02$ could suggest that nitrogen enrichment occurs on longer timescales than oxygen and iron. However, the observed $\sigma_{\textrm{N/H}}$ significantly exceeds the predictions of \cite{matthee2018}, and although the predictions are for scaling relations at $z=0.1$, it is not obvious why one would observe much higher scatter at $z\sim2$. If this increase is real, it could imply differences in the production of nitrogen in lower-metallicity galaxies, such as those at high redshift. Unfortunately, there are not significantly better observational constraints that can be placed on N/H, as [\ion{N}{2}]$\lambda\lambda6549,6585$ is the only nitrogen transition commonly observed in high-$z$ spectra. It is possible that in coming years, it will be possible to measure N/H directly in high-$z$ galaxies with observations of the auroral [\ion{N}{2}]$\lambda5755$ line with JWST; however, samples will be small owing to the extreme faintness of the auroral line relative to the nebular line.

\subsection{Constraints on Gas Flows}

Many studies use the effective yield $y_{\rm eff}\equiv Z/\ln(f_{\rm gas}^{-1})$, to understand the relationship between inflows and outflows and the shape of the O-MZR \citep[e.g.,][]{edmunds1990,garnett2002,dalcanton2007}. In a closed-box model, $y_{\rm eff} = y$, the true yield, which is assumed to be constant; differences in $y_{\rm eff}$ can be attributed to gas flows in and out of the galaxy. In $z\sim0$ galaxies, $y_{\rm eff}$ is seen to increase significantly with increasing baryonic mass \citep{tremonti2004}, which is interpreted as preferential metal loss from lower-mass galaxies due to their shallower potential wells. In contrast, high-$z$ galaxies show the opposite trend \citep{erb2006metal,mannucci2009,troncoso2014}, with $y_{\rm eff}$ \emph{decreasing} slightly with increasing mass. The implication of this result is that the efficiency of outflows does not decline with M$_{\ast}$ in $z\gtrsim2$ galaxies, and high-M$_{\ast}$ galaxies at these redshifts may even be more efficient at ejecting metals than their lower-M$_{\ast}$ counterparts \citep{torrey2019}. Indeed, having mass outflow rates scale proportionally with SFR naturally produces the decline in $y_{\rm eff}$ seen at high masses. Similarly, recent work by \citet{jian2018} suggests that the parallel slopes of the O-MZR and Fe-MZR at high-$z$ and relatively flatter O-MZR at $z\sim0$ are a consequence of larger metal loading factors at early times. The ability to compare the O-MZR and Fe-MZR at multiple redshifts is increasingly used in assessing the success of cosmological simulations \citep{ma2016mzr,derossi2017}.

Although we have not measured $f_{\rm gas}$ and $y_{\rm eff}$ for our sample, the shallower slope of the \texttt{GalDNA} O-MZR should translate to a steeper decline in $y_{\rm eff}$ with M$_{\ast}$ than seen in previous high-$z$ samples. This assumes the correlation between $f_{\rm gas}$ and M$_{\ast}$ for our galaxies is similar to that reported by \citet{erb2006metal}, a reasonable assumption, given that they share the same parent sample. A steep decline in $y_{\rm eff}$ requires massive galaxies to be even more efficient at ejecting metals from their ISM and implies very high mass-loading factors for outflows. This finding is more easily explained by a combination of enriched outflows \emph{and} unenriched inflows that scale with SFR \citep{erb2008}, which also produces a more marked decline in $y_{\rm eff}$ with M$_{\ast}$---and thus a shallower O-MZR slope.

Although disentangling the relative importance of inflows and outflows remains challenging, the CGM provides a ``fossil record" of past outflow activity in the form of heavy elements deposited there by galactic winds. The CGM contains a significant fraction of all the metals created in stars, and the distribution and physical state of this enriched gas can provide constraints on energetic feedback in galaxies \citep{tumlinson2011,peeples2014,oppenheimer2016,rudie2019}. Metal-bearing gaseous structures in the IGM and CGM at $z\sim2-3$ appear to be very small \citep[$\lesssim100$~pc; ][]{rauch2001,rudie2019}, and there are significant variations in the metal content and abundance pattern of the CGM of single galaxies \citep{zahedy2019,zahedy2021,cooper2021}, suggesting that metals are often poorly mixed in the gas around galaxies. Future studies comparing the composition of the CGM with other components of the galaxy, including the ISM and stars, will offer critical insight into the baryonic flows that profoundly shape the evolution of galaxies. 

\section{Summary}
\label{sec:summary}

We have inferred the chemical abundances of oxygen, nitrogen, and iron in 195 individual star-forming galaxies at $z\sim2-3$. Self-consistent measurements of O/H, N/H, and Fe/H were made using \texttt{GalDNA} (Section~\ref{sec:galdna}), which uses BPASS stellar population synthesis models together with Cloudy photoionization models to predict the rest-optical nebular spectra of galaxies. These results comprise the largest sample of high-$z$ galaxies with multiple elemental abundances to date and can be used to address questions related to the chemical evolution of galaxies in the early universe. Our key findings are that:

\begin{enumerate}
\item The photoionized gas and massive stars in $z\sim2$ galaxies have sub-solar abundances of O, N, and Fe, which increase with increasing M$_{\ast}$. The fiducial abundances at M$_{\ast}=10^{10}$~M$_{\odot}$ are 
\begin{eqnarray}
12+\log(\textrm{O/H}) = 8.35&\pm&0.02\textrm{  or  }45\%~ (\textrm{O/H})_{\odot}\nonumber \\
12+\log(\textrm{N/H}) = 7.07&\pm&0.03\textrm{  or  }17\%~ (\textrm{N/H})_{\odot}\nonumber \\
\textrm{[Fe/H]} = -0.69&\pm&0.02\textrm{  or  }20\%~ (\textrm{Fe/H})_{\odot}\nonumber
\end{eqnarray}

\item The O-MZR based on the \texttt{GalDNA} method is relatively shallow ($\beta_{\textrm{O/H}}=0.14\pm0.05$), with moderate intrinsic scatter ($\sigma_{\textrm{O/H}}=0.21\pm0.01$~dex, Figure~\ref{fig:omzr}). Although qualitatively similar to other O-MZRs in the literature, it is shallower and has larger scatter than relations based on strong-line O/H.

\item The slope of the O3N2 O-MZR (and other strong-line-based relations) depends on the coefficient(s) in the calibration (Figure~\ref{fig:omzr_slope}), which in turn depend on underlying correlations among abundances and other physical conditions like ionization parameter in the calibration sample. In other words, the correlation between $U$ and O/H in the \emph{calibration sample} affects the slope of the O-MZR for \emph{any other sample} for which that calibration is used. This can introduce systematic biases in physical interpretations of the form of such scaling relations.

\item The larger intrinsic scatter observed in the \texttt{GalDNA} O-MZR is the result of explicitly accounting for the ionization conditions in galaxies, which are strongly correlated with many line ratios and with M$_{\ast}$, but not as strongly correlated with O/H. Conversely, the strong correlation between $U$ and many commonly-used line ratios used to infer O/H can artificially suppress the scatter in strong-line-based O-MZRs (Figure~\ref{fig:omzr_intsc}).

\item The O-MZR and Fe-MZR (Figure~\ref{fig:femzr}) have nearly identical slopes across two decades in $M_{\ast}$, indicating a high level of $\alpha$-enhancement equivalent to $\textrm{O/Fe}\approx2.2\times(\textrm{O/Fe})_{\odot}$ even at relatively large M$_{\ast}$ (Figure~\ref{fig:mzrcomp}).

\item The intrinsic scatter in the O-MZR and Fe-MZR is very similar (Figure~\ref{fig:intsc}), which would be the case if the majority of the Fe had been produced along with the O by CCSNe, rather than by Type Ia SNe. The N-MZR has significantly larger scatter, potentially reflecting a longer timescale for nitrogen production.
\end{enumerate}
Our results demonstrate the potential to measure multiple elemental abundances in $z\sim2-3$ galaxies using emission lines that are commonly observed in their rest-optical spectra. Knowledge of galaxies abundance patterns---rather than just a single measure of their enrichment---should prove valuable in comparing galaxies' interstellar medium with their circumgalactic medium to understand gas flows in and out of galaxies. Moreover, abundance patterns can be used to link galaxy populations observed at different redshifts with one another by looking for commonalities in, for example, $\alpha$-enhancement.

Methods like \texttt{GalDNA} capable of accurately characterizing the slope \emph{and} scatter in abundance scaling relations for multiple elements can also be adapted for samples at other redshifts where different emission lines may be accessible (e.g., in the near-UV). Continuing to develop such methods will be key to extracting maximum scientific value from both existing galaxy surveys and observations from future facilities, including JWST, which will provide new opportunities to study some of the earliest galaxies in the universe.

\begin{acknowledgements}
The data presented in this paper were obtained at the W.M. Keck Observatory, which is operated as a scientific partnership among the California Institute of Technology, the University of California, and the National Aeronautics and Space Administration. The Observatory was made possible by the generous financial support of the W.M. Keck Foundation. This work has also been supported in part by the US National Science Foundation (NSF) through grants AST-0908805 and AST-1313472 (ALS and CCS). The authors wish to recognize and acknowledge the significant cultural role and reverence that the summit of Maunakea has within the indigenous Hawaiian community. We are privileged to have the opportunity to conduct observations from this mountain.
\end{acknowledgements}

\facilities{Keck:I (LRIS, MOSFIRE), Magellan:Baade (FourStar), Hale (WIRC), Spitzer (IRAC)}
\software{\texttt{GalDNA}, \texttt{LINMIX\_ERR} \citep{kelly2007}, BPASSv2 \citep{eldridge2016,stanway2016}}
\bibliography{strom_ref_library}

\begin{thebibliography}{}
\expandafter\ifx\csname natexlab\endcsname\relax\def\natexlab#1{#1}\fi
\providecommand{\url}[1]{\href{#1}{#1}}
\providecommand{\dodoi}[1]{doi:~\href{http://doi.org/#1}{\nolinkurl{#1}}}
\providecommand{\doeprint}[1]{\href{http://ascl.net/#1}{\nolinkurl{http://ascl.net/#1}}}
\providecommand{\doarXiv}[1]{\href{https://arxiv.org/abs/#1}{\nolinkurl{https://arxiv.org/abs/#1}}}

\bibitem[{{Adelberger} {et~al.}(2004){Adelberger}, {Steidel}, {Shapley},
  {Hunt}, {Erb}, {Reddy}, \& {Pettini}}]{adelberger2004}
{Adelberger}, K.~L., {Steidel}, C.~C., {Shapley}, A.~E., {et~al.} 2004, \apj,
  607, 226, \dodoi{10.1086/383221}

\bibitem[{{Aller}(1954)}]{aller1954}
{Aller}, L.~H. 1954, \apj, 120, 401, \dodoi{10.1086/145931}

\bibitem[{{Andrews} \& {Martini}(2013)}]{andrews2013}
{Andrews}, B.~H., \& {Martini}, P. 2013, \apj, 765, 140,
  \dodoi{10.1088/0004-637X/765/2/140}

\bibitem[{{Asplund} {et~al.}(2009){Asplund}, {Grevesse}, {Sauval}, \&
  {Scott}}]{asplund2009}
{Asplund}, M., {Grevesse}, N., {Sauval}, A.~J., \& {Scott}, P. 2009, \araa, 47,
  481, \dodoi{10.1146/annurev.astro.46.060407.145222}

\bibitem[{{Baldwin} {et~al.}(1981){Baldwin}, {Phillips}, \&
  {Terlevich}}]{baldwin1981}
{Baldwin}, J.~A., {Phillips}, M.~M., \& {Terlevich}, R. 1981, \pasp, 93, 5,
  \dodoi{10.1086/130766}

\bibitem[{{Barrera-Ballesteros} {et~al.}(2017){Barrera-Ballesteros},
  {S{\'a}nchez}, {Heckman}, {Blanc}, \& {MaNGA Team}}]{barrera-ballesteros2017}
{Barrera-Ballesteros}, J.~K., {S{\'a}nchez}, S.~F., {Heckman}, T., {Blanc},
  G.~A., \& {MaNGA Team}. 2017, \apj, 844, 80, \dodoi{10.3847/1538-4357/aa7aa9}

\bibitem[{{Berg} {et~al.}(2020){Berg}, {Pogge}, {Skillman}, {Croxall},
  {Moustakas}, {Rogers}, \& {Sun}}]{berg2020}
{Berg}, D.~A., {Pogge}, R.~W., {Skillman}, E.~D., {et~al.} 2020, \apj, 893, 96,
  \dodoi{10.3847/1538-4357/ab7eab}

\bibitem[{{Berg} {et~al.}(2012){Berg}, {Skillman}, {Marble}, {van Zee},
  {Engelbracht}, {Lee}, {Kennicutt}, {Calzetti}, {Dale}, \&
  {Johnson}}]{berg2012}
{Berg}, D.~A., {Skillman}, E.~D., {Marble}, A.~R., {et~al.} 2012, \apj, 754,
  98, \dodoi{10.1088/0004-637X/754/2/98}

\bibitem[{{Bian} {et~al.}(2018){Bian}, {Kewley}, \& {Dopita}}]{bian2018}
{Bian}, F., {Kewley}, L.~J., \& {Dopita}, M.~A. 2018, \apj, 859, 175,
  \dodoi{10.3847/1538-4357/aabd74}

\bibitem[{{Blanc} {et~al.}(2015){Blanc}, {Kewley}, {Vogt}, \&
  {Dopita}}]{blanc2015}
{Blanc}, G.~A., {Kewley}, L., {Vogt}, F.~P.~A., \& {Dopita}, M.~A. 2015, \apj,
  798, 99, \dodoi{10.1088/0004-637X/798/2/99}

\bibitem[{{Bothwell} {et~al.}(2013){Bothwell}, {Maiolino}, {Kennicutt},
  {Cresci}, {Mannucci}, {Marconi}, \& {Cicone}}]{bothwell2013}
{Bothwell}, M.~S., {Maiolino}, R., {Kennicutt}, R., {et~al.} 2013, \mnras, 433,
  1425, \dodoi{10.1093/mnras/stt817}

\bibitem[{{Brooks} {et~al.}(2007){Brooks}, {Governato}, {Booth}, {Willman},
  {Gardner}, {Wadsley}, {Stinson}, \& {Quinn}}]{brooks2007}
{Brooks}, A.~M., {Governato}, F., {Booth}, C.~M., {et~al.} 2007, \apjl, 655,
  L17, \dodoi{10.1086/511765}

\bibitem[{{Bruzual} \& {Charlot}(2003)}]{bruzual2003}
{Bruzual}, G., \& {Charlot}, S. 2003, \mnras, 344, 1000,
  \dodoi{10.1046/j.1365-8711.2003.06897.x}

\bibitem[{{Chabrier}(2003)}]{chabrier2003}
{Chabrier}, G. 2003, \pasp, 115, 763, \dodoi{10.1086/376392}

\bibitem[{{Conroy} {et~al.}(2014){Conroy}, {Graves}, \& {van
  Dokkum}}]{conroy2014}
{Conroy}, C., {Graves}, G.~J., \& {van Dokkum}, P.~G. 2014, \apj, 780, 33,
  \dodoi{10.1088/0004-637X/780/1/33}

\bibitem[{{Cooper} {et~al.}(2021){Cooper}, {Rudie}, {Chen}, {Johnson},
  {Zahedy}, {Chen}, {Boettcher}, {Walth}, {Cantalupo}, {Cooksey},
  {Faucher-Gigu{\`e}re}, {Greene}, {Lopez}, {Mulchaey}, {Penton}, {Petitjean},
  {Putman}, {Rafelski}, {Rauch}, {Schaye}, \& {Simcoe}}]{cooper2021}
{Cooper}, T.~J., {Rudie}, G.~C., {Chen}, H.-W., {et~al.} 2021, \mnras, 508,
  4359, \dodoi{10.1093/mnras/stab2869}

\bibitem[{{Cullen} {et~al.}(2019){Cullen}, {McLure}, {Dunlop}, {Khochfar},
  {Dav{\'e}}, {Amor{\'\i}n}, {Bolzonella}, {Carnall}, {Castellano}, {Cimatti},
  {Cirasuolo}, {Cresci}, {Fynbo}, {Fontanot}, {Gargiulo}, {Garilli}, {Guaita},
  {Hathi}, {Hibon}, {Mannucci}, {Marchi}, {McLeod}, {Pentericci}, {Pozzetti},
  {Shapley}, {Talia}, \& {Zamorani}}]{cullen2019}
{Cullen}, F., {McLure}, R.~J., {Dunlop}, J.~S., {et~al.} 2019, \mnras, 487,
  2038, \dodoi{10.1093/mnras/stz1402}

\bibitem[{{Cullen} {et~al.}(2021){Cullen}, {Shapley}, {McLure}, {Dunlop},
  {Sanders}, {Topping}, {Reddy}, {Amor{\'\i}n}, {Begley}, {Bolzonella},
  {Calabr{\'o}}, {Carnall}, {Castellano}, {Cimatti}, {Cirasuolo}, {Cresci},
  {Fontana}, {Fontanot}, {Garilli}, {Guaita}, {Hamadouche}, {Hathi},
  {Mannucci}, {McLeod}, {Pentericci}, {Saxena}, {Talia}, \&
  {Zamorani}}]{cullen2021}
{Cullen}, F., {Shapley}, A.~E., {McLure}, R.~J., {et~al.} 2021, \mnras,
  \dodoi{10.1093/mnras/stab1340}

\bibitem[{{Daddi} {et~al.}(2004){Daddi}, {Cimatti}, {Renzini}, {Fontana},
  {Mignoli}, {Pozzetti}, {Tozzi}, \& {Zamorani}}]{daddi2004}
{Daddi}, E., {Cimatti}, A., {Renzini}, A., {et~al.} 2004, \apj, 617, 746,
  \dodoi{10.1086/425569}

\bibitem[{{Dalcanton}(2007)}]{dalcanton2007}
{Dalcanton}, J.~J. 2007, \apj, 658, 941, \dodoi{10.1086/508913}

\bibitem[{{De Rossi} {et~al.}(2017){De Rossi}, {Bower}, {Font}, {Schaye}, \&
  {Theuns}}]{derossi2017}
{De Rossi}, M.~E., {Bower}, R.~G., {Font}, A.~S., {Schaye}, J., \& {Theuns}, T.
  2017, \mnras, 472, 3354, \dodoi{10.1093/mnras/stx2158}

\bibitem[{{Edmunds}(1990)}]{edmunds1990}
{Edmunds}, M.~G. 1990, \mnras, 246, 678

\bibitem[{{Eldridge} \& {Stanway}(2016)}]{eldridge2016}
{Eldridge}, J.~J., \& {Stanway}, E.~R. 2016, \mnras, 462, 3302,
  \dodoi{10.1093/mnras/stw1772}

\bibitem[{{Ellison} {et~al.}(2008){Ellison}, {Patton}, {Simard}, \&
  {McConnachie}}]{ellison2008}
{Ellison}, S.~L., {Patton}, D.~R., {Simard}, L., \& {McConnachie}, A.~W. 2008,
  \aj, 135, 1877, \dodoi{10.1088/0004-6256/135/5/1877}

\bibitem[{{Erb}(2008)}]{erb2008}
{Erb}, D.~K. 2008, \apj, 674, 151, \dodoi{10.1086/524727}

\bibitem[{{Erb} {et~al.}(2010){Erb}, {Pettini}, {Shapley}, {Steidel}, {Law}, \&
  {Reddy}}]{erb2010}
{Erb}, D.~K., {Pettini}, M., {Shapley}, A.~E., {et~al.} 2010, \apj, 719, 1168,
  \dodoi{10.1088/0004-637X/719/2/1168}

\bibitem[{{Erb} {et~al.}(2006{\natexlab{a}}){Erb}, {Shapley}, {Pettini},
  {Steidel}, {Reddy}, \& {Adelberger}}]{erb2006metal}
{Erb}, D.~K., {Shapley}, A.~E., {Pettini}, M., {et~al.} 2006{\natexlab{a}},
  \apj, 644, 813, \dodoi{10.1086/503623}

\bibitem[{{Erb} {et~al.}(2006{\natexlab{b}}){Erb}, {Steidel}, {Shapley},
  {Pettini}, {Reddy}, \& {Adelberger}}]{erb2006mass}
{Erb}, D.~K., {Steidel}, C.~C., {Shapley}, A.~E., {et~al.} 2006{\natexlab{b}},
  \apj, 646, 107, \dodoi{10.1086/504891}

\bibitem[{{Esteban} {et~al.}(2014){Esteban}, {Garc{\'{\i}}a-Rojas}, {Carigi},
  {Peimbert}, {Bresolin}, {L{\'o}pez-S{\'a}nchez}, \&
  {Mesa-Delgado}}]{esteban2014}
{Esteban}, C., {Garc{\'{\i}}a-Rojas}, J., {Carigi}, L., {et~al.} 2014, \mnras,
  443, 624, \dodoi{10.1093/mnras/stu1177}

\bibitem[{{Esteban} {et~al.}(2004){Esteban}, {Peimbert}, {Garc{\'{\i}}a-Rojas},
  {Ruiz}, {Peimbert}, \& {Rodr{\'{\i}}guez}}]{esteban2004}
{Esteban}, C., {Peimbert}, M., {Garc{\'{\i}}a-Rojas}, J., {et~al.} 2004,
  \mnras, 355, 229, \dodoi{10.1111/j.1365-2966.2004.08313.x}

\bibitem[{{Faber} {et~al.}(1992){Faber}, {Worthey}, \& {Gonzales}}]{faber1992}
{Faber}, S.~M., {Worthey}, G., \& {Gonzales}, J.~J. 1992, in The Stellar
  Populations of Galaxies, ed. B.~{Barbuy} \& A.~{Renzini}, Vol. 149, 255

\bibitem[{{Ferland} {et~al.}(2013){Ferland}, {Porter}, {van Hoof}, {Williams},
  {Abel}, {Lykins}, {Shaw}, {Henney}, \& {Stancil}}]{ferland2013}
{Ferland}, G.~J., {Porter}, R.~L., {van Hoof}, P.~A.~M., {et~al.} 2013, \rmxaa,
  49, 137

\bibitem[{{Franx} {et~al.}(2003){Franx}, {Labb{\'e}}, {Rudnick}, {van Dokkum},
  {Daddi}, {F{\"o}rster Schreiber}, {Moorwood}, {Rix}, {R{\"o}ttgering}, {van
  der Wel}, {van der Werf}, \& {van Starkenburg}}]{franx2003}
{Franx}, M., {Labb{\'e}}, I., {Rudnick}, G., {et~al.} 2003, \apjl, 587, L79,
  \dodoi{10.1086/375155}

\bibitem[{{Gallazzi} {et~al.}(2005){Gallazzi}, {Charlot}, {Brinchmann},
  {White}, \& {Tremonti}}]{gallazzi2005}
{Gallazzi}, A., {Charlot}, S., {Brinchmann}, J., {White}, S. D.~M., \&
  {Tremonti}, C.~A. 2005, \mnras, 362, 41,
  \dodoi{10.1111/j.1365-2966.2005.09321.x}

\bibitem[{{Gardner} {et~al.}(2006){Gardner}, {Mather}, {Clampin}, {Doyon},
  {Greenhouse}, {Hammel}, {Hutchings}, {Jakobsen}, {Lilly}, {Long}, {Lunine},
  {McCaughrean}, {Mountain}, {Nella}, {Rieke}, {Rieke}, {Rix}, {Smith},
  {Sonneborn}, {Stiavelli}, {Stockman}, {Windhorst}, \& {Wright}}]{gardner2006}
{Gardner}, J.~P., {Mather}, J.~C., {Clampin}, M., {et~al.} 2006, \ssr, 123,
  485, \dodoi{10.1007/s11214-006-8315-7}

\bibitem[{{Garnett}(2002)}]{garnett2002}
{Garnett}, D.~R. 2002, \apj, 581, 1019, \dodoi{10.1086/344301}

\bibitem[{{Gonz{\'a}lez Delgado} {et~al.}(2014){Gonz{\'a}lez Delgado}, {Cid
  Fernandes}, {Garc{\'\i}a-Benito}, {P{\'e}rez}, {de Amorim},
  {Cortijo-Ferrero}, {Lacerda}, {L{\'o}pez Fern{\'a}ndez}, {S{\'a}nchez}, {Vale
  Asari}, {Alves}, {Bland-Hawthorn}, {Galbany}, {Gallazzi}, {Husemann},
  {Bekeraite}, {Jungwiert}, {L{\'o}pez-S{\'a}nchez}, {de Lorenzo-C{\'a}ceres},
  {Marino}, {Mast}, {Moll{\'a}}, {del Olmo}, {S{\'a}nchez-Bl{\'a}zquez}, {van
  de Ven}, {V{\'\i}lchez}, {Walcher}, {Wisotzki}, {Ziegler}, \& {CALIFA
  Collaboration}}]{gonzalez-delgado2014}
{Gonz{\'a}lez Delgado}, R.~M., {Cid Fernandes}, R., {Garc{\'\i}a-Benito}, R.,
  {et~al.} 2014, \apjl, 791, L16, \dodoi{10.1088/2041-8205/791/1/L16}

\bibitem[{{Izotov} \& {Thuan}(1999)}]{izotov1999}
{Izotov}, Y.~I., \& {Thuan}, T.~X. 1999, \apj, 511, 639, \dodoi{10.1086/306708}

\bibitem[{{Jafariyazani} {et~al.}(2020){Jafariyazani}, {Newman}, {Mobasher},
  {Belli}, {Ellis}, \& {Patel}}]{jafariyazani2020}
{Jafariyazani}, M., {Newman}, A.~B., {Mobasher}, B., {et~al.} 2020, \apjl, 897,
  L42, \dodoi{10.3847/2041-8213/aba11c}

\bibitem[{{Jeong} {et~al.}(2020){Jeong}, {Shapley}, {Sanders}, {Runco},
  {Topping}, {Reddy}, {Kriek}, {Coil}, {Mobasher}, {Siana}, {Shivaei},
  {Freeman}, {Azadi}, {Price}, {Leung}, {Fetherolf}, {de Groot}, {Zick},
  {Fornasini}, \& {Barro}}]{jeong2020}
{Jeong}, M.-S., {Shapley}, A.~E., {Sanders}, R.~L., {et~al.} 2020, \apjl, 902,
  L16, \dodoi{10.3847/2041-8213/abba7a}

\bibitem[{{Juneau} {et~al.}(2011){Juneau}, {Dickinson}, {Alexander}, \&
  {Salim}}]{juneau2011}
{Juneau}, S., {Dickinson}, M., {Alexander}, D.~M., \& {Salim}, S. 2011, \apj,
  736, 104, \dodoi{10.1088/0004-637X/736/2/104}

\bibitem[{{Juneau} {et~al.}(2014){Juneau}, {Bournaud}, {Charlot}, {Daddi},
  {Elbaz}, {Trump}, {Brinchmann}, {Dickinson}, {Duc}, {Gobat}, {Jean-Baptiste},
  {Le Floc'h}, {Lehnert}, {Pacifici}, {Pannella}, \& {Schreiber}}]{juneau2014}
{Juneau}, S., {Bournaud}, F., {Charlot}, S., {et~al.} 2014, \apj, 788, 88,
  \dodoi{10.1088/0004-637X/788/1/88}

\bibitem[{{Kaasinen} {et~al.}(2018){Kaasinen}, {Kewley}, {Bian}, {Groves},
  {Kashino}, {Silverman}, \& {Kartaltepe}}]{kaasinen2018}
{Kaasinen}, M., {Kewley}, L., {Bian}, F., {et~al.} 2018, \mnras, 477, 5568,
  \dodoi{10.1093/mnras/sty1012}

\bibitem[{{Kelly}(2007)}]{kelly2007}
{Kelly}, B.~C. 2007, \apj, 665, 1489, \dodoi{10.1086/519947}

\bibitem[{{Kewley} \& {Ellison}(2008)}]{kewley2008}
{Kewley}, L.~J., \& {Ellison}, S.~L. 2008, \apj, 681, 1183,
  \dodoi{10.1086/587500}

\bibitem[{{Kojima} {et~al.}(2017){Kojima}, {Ouchi}, {Nakajima}, {Shibuya},
  {Harikane}, \& {Ono}}]{kojima2017}
{Kojima}, T., {Ouchi}, M., {Nakajima}, K., {et~al.} 2017, \pasj, 69, 44,
  \dodoi{10.1093/pasj/psx017}

\bibitem[{{Kriek} {et~al.}(2016){Kriek}, {Conroy}, {van Dokkum}, {Shapley},
  {Choi}, {Reddy}, {Siana}, {van de Voort}, {Coil}, \& {Mobasher}}]{kriek2016}
{Kriek}, M., {Conroy}, C., {van Dokkum}, P.~G., {et~al.} 2016, \nat, 540, 248,
  \dodoi{10.1038/nature20570}

\bibitem[{{Kriek} {et~al.}(2019){Kriek}, {Price}, {Conroy}, {Suess}, {Mowla},
  {Pasha}, {Bezanson}, {van Dokkum}, \& {Barro}}]{kriek2019}
{Kriek}, M., {Price}, S.~H., {Conroy}, C., {et~al.} 2019, \apjl, 880, L31,
  \dodoi{10.3847/2041-8213/ab2e75}

\bibitem[{{Lara-L{\'o}pez} {et~al.}(2010){Lara-L{\'o}pez}, {Cepa},
  {Bongiovanni}, {P{\'e}rez Garc{\'\i}a}, {Ederoclite}, {Casta{\~n}eda},
  {Fern{\'a}ndez Lorenzo}, {Povi{\'c}}, \&
  {S{\'a}nchez-Portal}}]{lara-lopez2010}
{Lara-L{\'o}pez}, M.~A., {Cepa}, J., {Bongiovanni}, A., {et~al.} 2010, \aap,
  521, L53, \dodoi{10.1051/0004-6361/201014803}

\bibitem[{{Leitherer} {et~al.}(2010){Leitherer}, {Ortiz Ot{\'a}lvaro},
  {Bresolin}, {Kudritzki}, {Lo Faro}, {Pauldrach}, {Pettini}, \&
  {Rix}}]{leitherer2010}
{Leitherer}, C., {Ortiz Ot{\'a}lvaro}, P.~A., {Bresolin}, F., {et~al.} 2010,
  \apjs, 189, 309, \dodoi{10.1088/0067-0049/189/2/309}

\bibitem[{{Lequeux} {et~al.}(1979){Lequeux}, {Peimbert}, {Rayo}, {Serrano}, \&
  {Torres-Peimbert}}]{lequeux1979}
{Lequeux}, J., {Peimbert}, M., {Rayo}, J.~F., {Serrano}, A., \&
  {Torres-Peimbert}, S. 1979, \aap, 500, 145

\bibitem[{{Lian} {et~al.}(2018){Lian}, {Thomas}, \& {Maraston}}]{jian2018}
{Lian}, J., {Thomas}, D., \& {Maraston}, C. 2018, \mnras, 481, 4000,
  \dodoi{10.1093/mnras/sty2506}

\bibitem[{{Ma} {et~al.}(2016){Ma}, {Hopkins}, {Faucher-Gigu{\`e}re}, {Zolman},
  {Muratov}, {Kere{\v{s}}}, \& {Quataert}}]{ma2016mzr}
{Ma}, X., {Hopkins}, P.~F., {Faucher-Gigu{\`e}re}, C.-A., {et~al.} 2016,
  \mnras, 456, 2140, \dodoi{10.1093/mnras/stv2659}

\bibitem[{{Maiolino} \& {Mannucci}(2019)}]{maiolino2019}
{Maiolino}, R., \& {Mannucci}, F. 2019, \aapr, 27, 3,
  \dodoi{10.1007/s00159-018-0112-2}

\bibitem[{{Maiolino} {et~al.}(2008){Maiolino}, {Nagao}, {Grazian}, {Cocchia},
  {Marconi}, {Mannucci}, {Cimatti}, {Pipino}, {Ballero}, {Calura}, {Chiappini},
  {Fontana}, {Granato}, {Matteucci}, {Pastorini}, {Pentericci}, {Risaliti},
  {Salvati}, \& {Silva}}]{maiolino2008}
{Maiolino}, R., {Nagao}, T., {Grazian}, A., {et~al.} 2008, \aap, 488, 463,
  \dodoi{10.1051/0004-6361:200809678}

\bibitem[{{Mannucci} {et~al.}(2010){Mannucci}, {Cresci}, {Maiolino}, {Marconi},
  \& {Gnerucci}}]{mannucci2010}
{Mannucci}, F., {Cresci}, G., {Maiolino}, R., {Marconi}, A., \& {Gnerucci}, A.
  2010, \mnras, 408, 2115, \dodoi{10.1111/j.1365-2966.2010.17291.x}

\bibitem[{{Mannucci} {et~al.}(2009){Mannucci}, {Cresci}, {Maiolino}, {Marconi},
  {Pastorini}, {Pozzetti}, {Gnerucci}, {Risaliti}, {Schneider}, {Lehnert}, \&
  {Salvati}}]{mannucci2009}
{Mannucci}, F., {Cresci}, G., {Maiolino}, R., {et~al.} 2009, \mnras, 398, 1915,
  \dodoi{10.1111/j.1365-2966.2009.15185.x}

\bibitem[{{Marino} {et~al.}(2013){Marino}, {Rosales-Ortega}, {S{\'a}nchez},
  {Gil de Paz}, {V{\'\i}lchez}, {Miralles-Caballero}, {Kehrig},
  {P{\'e}rez-Montero}, {Stanishev}, {Iglesias-P{\'a}ramo}, {D{\'\i}az},
  {Castillo-Morales}, {Kennicutt}, {L{\'o}pez-S{\'a}nchez}, {Galbany},
  {Garc{\'\i}a-Benito}, {Mast}, {Mendez-Abreu}, {Monreal-Ibero}, {Husemann},
  {Walcher}, {Garc{\'\i}a-Lorenzo}, {Masegosa}, {Del Olmo Orozco},
  {Mour{\~a}o}, {Ziegler}, {Moll{\'a}}, {Papaderos},
  {S{\'a}nchez-Bl{\'a}zquez}, {Gonz{\'a}lez Delgado}, {Falc{\'o}n-Barroso},
  {Roth}, {van de Ven}, \& {Califa Team}}]{marino2013}
{Marino}, R.~A., {Rosales-Ortega}, F.~F., {S{\'a}nchez}, S.~F., {et~al.} 2013,
  \aap, 559, A114, \dodoi{10.1051/0004-6361/201321956}

\bibitem[{{Matteucci} {et~al.}(2009){Matteucci}, {Spitoni}, {Recchi}, \&
  {Valiante}}]{matteucci2009}
{Matteucci}, F., {Spitoni}, E., {Recchi}, S., \& {Valiante}, R. 2009, \aap,
  501, 531, \dodoi{10.1051/0004-6361/200911869}

\bibitem[{{Matthee} \& {Schaye}(2018)}]{matthee2018}
{Matthee}, J., \& {Schaye}, J. 2018, \mnras, 479, L34,
  \dodoi{10.1093/mnrasl/sly093}

\bibitem[{{McGaugh}(1991)}]{mcgaugh1991}
{McGaugh}, S.~S. 1991, \apj, 380, 140, \dodoi{10.1086/170569}

\bibitem[{{McLean} {et~al.}(2012){McLean}, {Steidel}, {Epps}, {Konidaris},
  {Matthews}, {Adkins}, {Aliado}, {Brims}, {Canfield}, {Cromer}, {Fucik},
  {Kulas}, {Mace}, {Magnone}, {Rodriguez}, {Rudie}, {Trainor}, {Wang}, {Weber},
  \& {Weiss}}]{mclean2012}
{McLean}, I.~S., {Steidel}, C.~C., {Epps}, H.~W., {et~al.} 2012, in SPIE
  Conference Series, Vol. 8446, \dodoi{10.1117/12.924794}

\bibitem[{{Nomoto} {et~al.}(2006){Nomoto}, {Tominaga}, {Umeda}, {Kobayashi}, \&
  {Maeda}}]{nomoto2006}
{Nomoto}, K., {Tominaga}, N., {Umeda}, H., {Kobayashi}, C., \& {Maeda}, K.
  2006, Nuclear Physics A, 777, 424, \dodoi{10.1016/j.nuclphysa.2006.05.008}

\bibitem[{{Onodera} {et~al.}(2016){Onodera}, {Carollo}, {Lilly}, {Renzini},
  {Arimoto}, {Capak}, {Daddi}, {Scoville}, {Tacchella}, {Tatehora}, \&
  {Zamorani}}]{onodera2016}
{Onodera}, M., {Carollo}, C.~M., {Lilly}, S., {et~al.} 2016, \apj, 822, 42,
  \dodoi{10.3847/0004-637X/822/1/42}

\bibitem[{{Oppenheimer} {et~al.}(2016){Oppenheimer}, {Crain}, {Schaye},
  {Rahmati}, {Richings}, {Trayford}, {Tumlinson}, {Bower}, {Schaller}, \&
  {Theuns}}]{oppenheimer2016}
{Oppenheimer}, B.~D., {Crain}, R.~A., {Schaye}, J., {et~al.} 2016, \mnras, 460,
  2157, \dodoi{10.1093/mnras/stw1066}

\bibitem[{{Peeples} {et~al.}(2014){Peeples}, {Werk}, {Tumlinson},
  {Oppenheimer}, {Prochaska}, {Katz}, \& {Weinberg}}]{peeples2014}
{Peeples}, M.~S., {Werk}, J.~K., {Tumlinson}, J., {et~al.} 2014, \apj, 786, 54,
  \dodoi{10.1088/0004-637X/786/1/54}

\bibitem[{{P{\'e}rez-Montero}(2014)}]{perez-montero2014}
{P{\'e}rez-Montero}, E. 2014, \mnras, 441, 2663, \dodoi{10.1093/mnras/stu753}

\bibitem[{{Petitjean} {et~al.}(2008){Petitjean}, {Ledoux}, \&
  {Srianand}}]{petitjean2008}
{Petitjean}, P., {Ledoux}, C., \& {Srianand}, R. 2008, \aap, 480, 349,
  \dodoi{10.1051/0004-6361:20078607}

\bibitem[{{Pettini} {et~al.}(2002){Pettini}, {Ellison}, {Bergeron}, \&
  {Petitjean}}]{pettini2002}
{Pettini}, M., {Ellison}, S.~L., {Bergeron}, J., \& {Petitjean}, P. 2002, \aap,
  391, 21, \dodoi{10.1051/0004-6361:20020809}

\bibitem[{{Pettini} {et~al.}(1995){Pettini}, {Lipman}, \&
  {Hunstead}}]{pettini1995}
{Pettini}, M., {Lipman}, K., \& {Hunstead}, R.~W. 1995, \apj, 451, 100,
  \dodoi{10.1086/176203}

\bibitem[{{Pettini} \& {Pagel}(2004)}]{pettini2004}
{Pettini}, M., \& {Pagel}, B.~E.~J. 2004, \mnras, 348, L59,
  \dodoi{10.1111/j.1365-2966.2004.07591.x}

\bibitem[{{Pettini} {et~al.}(2008){Pettini}, {Zych}, {Steidel}, \&
  {Chaffee}}]{pettini2008}
{Pettini}, M., {Zych}, B.~J., {Steidel}, C.~C., \& {Chaffee}, F.~H. 2008,
  \mnras, 385, 2011, \dodoi{10.1111/j.1365-2966.2008.12951.x}

\bibitem[{{Pilyugin} {et~al.}(2012){Pilyugin}, {Grebel}, \&
  {Mattsson}}]{pilyugin2012}
{Pilyugin}, L.~S., {Grebel}, E.~K., \& {Mattsson}, L. 2012, \mnras, 424, 2316,
  \dodoi{10.1111/j.1365-2966.2012.21398.x}

\bibitem[{{Pilyugin} \& {Thuan}(2005)}]{pilyugin2005}
{Pilyugin}, L.~S., \& {Thuan}, T.~X. 2005, \apj, 631, 231,
  \dodoi{10.1086/432408}

\bibitem[{{Rauch} {et~al.}(2001){Rauch}, {Sargent}, \& {Barlow}}]{rauch2001}
{Rauch}, M., {Sargent}, W. L.~W., \& {Barlow}, T.~A. 2001, \apj, 554, 823,
  \dodoi{10.1086/321402}

\bibitem[{{Reddy} {et~al.}(2005){Reddy}, {Erb}, {Steidel}, {Shapley},
  {Adelberger}, \& {Pettini}}]{reddy2005}
{Reddy}, N.~A., {Erb}, D.~K., {Steidel}, C.~C., {et~al.} 2005, \apj, 633, 748,
  \dodoi{10.1086/444588}

\bibitem[{{Reddy} {et~al.}(2012){Reddy}, {Pettini}, {Steidel}, {Shapley},
  {Erb}, \& {Law}}]{reddy2012}
{Reddy}, N.~A., {Pettini}, M., {Steidel}, C.~C., {et~al.} 2012, \apj, 754, 25,
  \dodoi{10.1088/0004-637X/754/1/25}

\bibitem[{{Rudie} {et~al.}(2019){Rudie}, {Steidel}, {Pettini}, {Trainor},
  {Strom}, {Hummels}, {Reddy}, \& {Shapley}}]{rudie2019}
{Rudie}, G.~C., {Steidel}, C.~C., {Pettini}, M., {et~al.} 2019, \apj, 885, 61,
  \dodoi{10.3847/1538-4357/ab4255}

\bibitem[{{Rudie} {et~al.}(2012){Rudie}, {Steidel}, {Trainor}, {Rakic},
  {Bogosavljevi{\'c}}, {Pettini}, {Reddy}, {Shapley}, {Erb}, \&
  {Law}}]{rudie2012}
{Rudie}, G.~C., {Steidel}, C.~C., {Trainor}, R.~F., {et~al.} 2012, \apj, 750,
  67, \dodoi{10.1088/0004-637X/750/1/67}

\bibitem[{{S{\'a}nchez} {et~al.}(2012){S{\'a}nchez}, {Kennicutt}, {Gil de Paz},
  {van de Ven}, {V{\'{\i}}lchez}, {Wisotzki}, {Walcher}, {Mast}, {Aguerri},
  {Albiol-P{\'e}rez}, {Alonso-Herrero}, {Alves}, {Bakos}, {Bart{\'a}kov{\'a}},
  {Bland-Hawthorn}, {Boselli}, {Bomans}, {Castillo-Morales}, {Cortijo-Ferrero},
  {de Lorenzo-C{\'a}ceres}, {Del Olmo}, {Dettmar}, {D{\'{\i}}az}, {Ellis},
  {Falc{\'o}n-Barroso}, {Flores}, {Gallazzi}, {Garc{\'{\i}}a-Lorenzo},
  {Gonz{\'a}lez Delgado}, {Gruel}, {Haines}, {Hao}, {Husemann},
  {Igl{\'e}sias-P{\'a}ramo}, {Jahnke}, {Johnson}, {Jungwiert}, {Kalinova},
  {Kehrig}, {Kupko}, {L{\'o}pez-S{\'a}nchez}, {Lyubenova}, {Marino},
  {M{\'a}rmol-Queralt{\'o}}, {M{\'a}rquez}, {Masegosa}, {Meidt},
  {Mendez-Abreu}, {Monreal-Ibero}, {Montijo}, {Mour{\~a}o}, {Palacios-Navarro},
  {Papaderos}, {Pasquali}, {Peletier}, {P{\'e}rez}, {P{\'e}rez}, {Quirrenbach},
  {Rela{\~n}o}, {Rosales-Ortega}, {Roth}, {Ruiz-Lara},
  {S{\'a}nchez-Bl{\'a}zquez}, {Sengupta}, {Singh}, {Stanishev}, {Trager},
  {Vazdekis}, {Viironen}, {Wild}, {Zibetti}, \& {Ziegler}}]{sanchez2012}
{S{\'a}nchez}, S.~F., {Kennicutt}, R.~C., {Gil de Paz}, A., {et~al.} 2012,
  \aap, 538, A8, \dodoi{10.1051/0004-6361/201117353}

\bibitem[{{S{\'a}nchez} {et~al.}(2019){S{\'a}nchez}, {Barrera-Ballesteros},
  {L{\'o}pez-Cob{\'a}}, {Brough}, {Bryant}, {Bland-Hawthorn}, {Croom}, {van de
  Sande}, {Cortese}, {Goodwin}, {Lawrence}, {L{\'o}pez-S{\'a}nchez}, {Sweet},
  {Owers}, {Richards}, \& {Walcher}}]{sanchez2019}
{S{\'a}nchez}, S.~F., {Barrera-Ballesteros}, J.~K., {L{\'o}pez-Cob{\'a}}, C.,
  {et~al.} 2019, \mnras, 484, 3042, \dodoi{10.1093/mnras/stz019}

\bibitem[{{Sanders} {et~al.}(2017){Sanders}, {Shapley}, {Zhang}, \&
  {Yan}}]{sanders2017}
{Sanders}, R.~L., {Shapley}, A.~E., {Zhang}, K., \& {Yan}, R. 2017, \apj, 850,
  136, \dodoi{10.3847/1538-4357/aa93e4}

\bibitem[{{Sanders} {et~al.}(2015){Sanders}, {Shapley}, {Kriek}, {Reddy},
  {Freeman}, {Coil}, {Siana}, {Mobasher}, {Shivaei}, {Price}, \& {de
  Groot}}]{sanders2015}
{Sanders}, R.~L., {Shapley}, A.~E., {Kriek}, M., {et~al.} 2015, \apj, 799, 138,
  \dodoi{10.1088/0004-637X/799/2/138}

\bibitem[{{Sanders} {et~al.}(2016){Sanders}, {Shapley}, {Kriek}, {Reddy},
  {Freeman}, {Coil}, {Siana}, {Mobasher}, {Shivaei}, {Price}, \& {de
  Groot}}]{sanders2016}
---. 2016, \apj, 816, 23, \dodoi{10.3847/0004-637X/816/1/23}

\bibitem[{{Sanders} {et~al.}(2018){Sanders}, {Shapley}, {Kriek}, {Freeman},
  {Reddy}, {Siana}, {Coil}, {Mobasher}, {Dav{\'e}}, {Shivaei}, {Azadi},
  {Price}, {Leung}, {Fetherolf}, {de Groot}, {Zick}, {Fornasini}, \&
  {Barro}}]{sanders2018}
---. 2018, \apj, 858, 99, \dodoi{10.3847/1538-4357/aabcbd}

\bibitem[{{Sanders} {et~al.}(2020{\natexlab{a}}){Sanders}, {Shapley}, {Reddy},
  {Kriek}, {Siana}, {Coil}, {Mobasher}, {Shivaei}, {Freeman}, {Azadi}, {Price},
  {Leung}, {Fetherolf}, {de Groot}, {Zick}, {Fornasini}, \&
  {Barro}}]{sanders2020}
{Sanders}, R.~L., {Shapley}, A.~E., {Reddy}, N.~A., {et~al.}
  2020{\natexlab{a}}, \mnras, 491, 1427, \dodoi{10.1093/mnras/stz3032}

\bibitem[{{Sanders} {et~al.}(2020{\natexlab{b}}){Sanders}, {Jones}, {Shapley},
  {Reddy}, {Kriek}, {Coil}, {Siana}, {Mobasher}, {Shivaei}, {Price}, {Freeman},
  {Azadi}, {Leung}, {Fetherolf}, {Zick}, {de Groot}, {Barro}, \&
  {Fornasini}}]{sanders2020sulfur}
{Sanders}, R.~L., {Jones}, T., {Shapley}, A.~E., {et~al.} 2020{\natexlab{b}},
  \apjl, 888, L11, \dodoi{10.3847/2041-8213/ab5d40}

\bibitem[{{Sanders} {et~al.}(2021){Sanders}, {Shapley}, {Jones}, {Reddy},
  {Kriek}, {Siana}, {Coil}, {Mobasher}, {Shivaei}, {Dav{\'e}}, {Azadi},
  {Price}, {Leung}, {Freeman}, {Fetherolf}, {de Groot}, {Zick}, \&
  {Barro}}]{sanders2021}
{Sanders}, R.~L., {Shapley}, A.~E., {Jones}, T., {et~al.} 2021, \apj, 914, 19,
  \dodoi{10.3847/1538-4357/abf4c1}

\bibitem[{{Scannapieco} \& {Bildsten}(2005)}]{scannapieco2005}
{Scannapieco}, E., \& {Bildsten}, L. 2005, \apjl, 629, L85,
  \dodoi{10.1086/452632}

\bibitem[{{Segers} {et~al.}(2016){Segers}, {Schaye}, {Bower}, {Crain},
  {Schaller}, \& {Theuns}}]{segers2016}
{Segers}, M.~C., {Schaye}, J., {Bower}, R.~G., {et~al.} 2016, \mnras, 461,
  L102, \dodoi{10.1093/mnrasl/slw111}

\bibitem[{{Shapley} {et~al.}(2005){Shapley}, {Steidel}, {Erb}, {Reddy},
  {Adelberger}, {Pettini}, {Barmby}, \& {Huang}}]{shapley2005}
{Shapley}, A.~E., {Steidel}, C.~C., {Erb}, D.~K., {et~al.} 2005, \apj, 626,
  698, \dodoi{10.1086/429990}

\bibitem[{{Stanway} {et~al.}(2016){Stanway}, {Eldridge}, \&
  {Becker}}]{stanway2016}
{Stanway}, E.~R., {Eldridge}, J.~J., \& {Becker}, G.~D. 2016, \mnras, 456, 485,
  \dodoi{10.1093/mnras/stv2661}

\bibitem[{{Steidel} {et~al.}(2003){Steidel}, {Adelberger}, {Shapley},
  {Pettini}, {Dickinson}, \& {Giavalisco}}]{steidel2003}
{Steidel}, C.~C., {Adelberger}, K.~L., {Shapley}, A.~E., {et~al.} 2003, \apj,
  592, 728, \dodoi{10.1086/375772}

\bibitem[{{Steidel} {et~al.}(2004){Steidel}, {Shapley}, {Pettini},
  {Adelberger}, {Erb}, {Reddy}, \& {Hunt}}]{steidel2004}
{Steidel}, C.~C., {Shapley}, A.~E., {Pettini}, M., {et~al.} 2004, \apj, 604,
  534, \dodoi{10.1086/381960}

\bibitem[{{Steidel} {et~al.}(2016){Steidel}, {Strom}, {Pettini}, {Rudie},
  {Reddy}, \& {Trainor}}]{steidel2016}
{Steidel}, C.~C., {Strom}, A.~L., {Pettini}, M., {et~al.} 2016, \apj, 826, 159,
  \dodoi{10.3847/0004-637X/826/2/159}

\bibitem[{{Steidel} {et~al.}(2014){Steidel}, {Rudie}, {Strom}, {Pettini},
  {Reddy}, {Shapley}, {Trainor}, {Erb}, {Turner}, {Konidaris}, {Kulas}, {Mace},
  {Matthews}, \& {McLean}}]{steidel2014}
{Steidel}, C.~C., {Rudie}, G.~C., {Strom}, A.~L., {et~al.} 2014, \apj, 795,
  165, \dodoi{10.1088/0004-637X/795/2/165}

\bibitem[{{Strom} {et~al.}(2018){Strom}, {Steidel}, {Rudie}, {Trainor}, \&
  {Pettini}}]{strom2018}
{Strom}, A.~L., {Steidel}, C.~C., {Rudie}, G.~C., {Trainor}, R.~F., \&
  {Pettini}, M. 2018, \apj, 868, 117, \dodoi{10.3847/1538-4357/aae1a5}

\bibitem[{{Strom} {et~al.}(2017){Strom}, {Steidel}, {Rudie}, {Trainor},
  {Pettini}, \& {Reddy}}]{strom2017}
{Strom}, A.~L., {Steidel}, C.~C., {Rudie}, G.~C., {et~al.} 2017, \apj, 836,
  164, \dodoi{10.3847/1538-4357/836/2/164}

\bibitem[{{Theios} {et~al.}(2019){Theios}, {Steidel}, {Strom}, {Rudie},
  {Trainor}, \& {Reddy}}]{theios2019}
{Theios}, R.~L., {Steidel}, C.~C., {Strom}, A.~L., {et~al.} 2019, \apj, 871,
  128, \dodoi{10.3847/1538-4357/aaf386}

\bibitem[{{Thomas} {et~al.}(2018){Thomas}, {Dopita}, {Kewley}, {Groves},
  {Sutherland}, {Hopkins}, \& {Blanc}}]{thomas2018}
{Thomas}, A.~D., {Dopita}, M.~A., {Kewley}, L.~J., {et~al.} 2018, \apj, 856,
  89, \dodoi{10.3847/1538-4357/aab3db}

\bibitem[{{Thomas} {et~al.}(2010){Thomas}, {Maraston}, {Schawinski}, {Sarzi},
  \& {Silk}}]{thomas2010}
{Thomas}, D., {Maraston}, C., {Schawinski}, K., {Sarzi}, M., \& {Silk}, J.
  2010, \mnras, 404, 1775, \dodoi{10.1111/j.1365-2966.2010.16427.x}

\bibitem[{{Tinsley}(1979)}]{tinsley1979}
{Tinsley}, B.~M. 1979, \apj, 229, 1046, \dodoi{10.1086/157039}

\bibitem[{{Topping} {et~al.}(2020){Topping}, {Shapley}, {Reddy}, {Sanders},
  {Coil}, {Kriek}, {Mobasher}, \& {Siana}}]{topping2020}
{Topping}, M.~W., {Shapley}, A.~E., {Reddy}, N.~A., {et~al.} 2020, \mnras, 499,
  1652, \dodoi{10.1093/mnras/staa2941}

\bibitem[{{Torrey} {et~al.}(2019){Torrey}, {Vogelsberger}, {Marinacci},
  {Pakmor}, {Springel}, {Nelson}, {Naiman}, {Pillepich}, {Genel}, {Weinberger},
  \& {Hernquist}}]{torrey2019}
{Torrey}, P., {Vogelsberger}, M., {Marinacci}, F., {et~al.} 2019, \mnras, 484,
  5587, \dodoi{10.1093/mnras/stz243}

\bibitem[{{Tremonti} {et~al.}(2004){Tremonti}, {Heckman}, {Kauffmann},
  {Brinchmann}, {Charlot}, {White}, {Seibert}, {Peng}, {Schlegel}, {Uomoto},
  {Fukugita}, \& {Brinkmann}}]{tremonti2004}
{Tremonti}, C.~A., {Heckman}, T.~M., {Kauffmann}, G., {et~al.} 2004, \apj, 613,
  898, \dodoi{10.1086/423264}

\bibitem[{{Troncoso} {et~al.}(2014){Troncoso}, {Maiolino}, {Sommariva},
  {Cresci}, {Mannucci}, {Marconi}, {Meneghetti}, {Grazian}, {Cimatti},
  {Fontana}, {Nagao}, \& {Pentericci}}]{troncoso2014}
{Troncoso}, P., {Maiolino}, R., {Sommariva}, V., {et~al.} 2014, \aap, 563, A58,
  \dodoi{10.1051/0004-6361/201322099}

\bibitem[{{Tumlinson} {et~al.}(2017){Tumlinson}, {Peeples}, \&
  {Werk}}]{tumlinson2017}
{Tumlinson}, J., {Peeples}, M.~S., \& {Werk}, J.~K. 2017, \araa, 55, 389,
  \dodoi{10.1146/annurev-astro-091916-055240}

\bibitem[{{Tumlinson} {et~al.}(2011){Tumlinson}, {Thom}, {Werk}, {Prochaska},
  {Tripp}, {Weinberg}, {Peeples}, {O'Meara}, {Oppenheimer}, {Meiring}, {Katz},
  {Dav{\'e}}, {Ford}, \& {Sembach}}]{tumlinson2011}
{Tumlinson}, J., {Thom}, C., {Werk}, J.~K., {et~al.} 2011, Science, 334, 948,
  \dodoi{10.1126/science.1209840}

\bibitem[{{Vale Asari} {et~al.}(2016){Vale Asari}, {Stasi{\'n}ska}, {Morisset},
  \& {Cid Fernandes}}]{valeasari2016}
{Vale Asari}, N., {Stasi{\'n}ska}, G., {Morisset}, C., \& {Cid Fernandes}, R.
  2016, \mnras, 460, 1739, \dodoi{10.1093/mnras/stw971}

\bibitem[{{van Loon} {et~al.}(2021){van Loon}, {Mitchell}, \&
  {Schaye}}]{vanloon2021}
{van Loon}, M.~L., {Mitchell}, P.~D., \& {Schaye}, J. 2021, \mnras, 504, 4817,
  \dodoi{10.1093/mnras/stab1254}

\bibitem[{{van Zee} {et~al.}(1998){van Zee}, {Salzer}, \&
  {Haynes}}]{vanzee1998}
{van Zee}, L., {Salzer}, J.~J., \& {Haynes}, M.~P. 1998, \apjl, 497, L1,
  \dodoi{10.1086/311263}

\bibitem[{{Veilleux} \& {Osterbrock}(1987)}]{veilleux1987}
{Veilleux}, S., \& {Osterbrock}, D.~E. 1987, \apjs, 63, 295,
  \dodoi{10.1086/191166}

\bibitem[{{Yabe} {et~al.}(2014){Yabe}, {Ohta}, {Iwamuro}, {Akiyama}, {Tamura},
  {Yuma}, {Kimura}, {Takato}, {Moritani}, {Sumiyoshi}, {Maihara}, {Silverman},
  {Dalton}, {Lewis}, {Bonfield}, {Lee}, {Curtis-Lake}, {Macaulay}, \&
  {Clarke}}]{yabe2014}
{Yabe}, K., {Ohta}, K., {Iwamuro}, F., {et~al.} 2014, \mnras, 437, 3647,
  \dodoi{10.1093/mnras/stt2185}

\bibitem[{{Yates} {et~al.}(2013){Yates}, {Henriques}, {Thomas}, {Kauffmann},
  {Johansson}, \& {White}}]{yates2013}
{Yates}, R.~M., {Henriques}, B., {Thomas}, P.~A., {et~al.} 2013, \mnras, 435,
  3500, \dodoi{10.1093/mnras/stt1542}

\bibitem[{{York} {et~al.}(2000){York}, {Adelman}, {Anderson}, {Anderson},
  {Annis}, {Bahcall}, {Bakken}, {Barkhouser}, {Bastian}, {Berman}, {Boroski},
  {Bracker}, {Briegel}, {Briggs}, {Brinkmann}, {Brunner}, {Burles}, {Carey},
  {Carr}, {Castander}, {Chen}, {Colestock}, {Connolly}, {Crocker}, {Csabai},
  {Czarapata}, {Davis}, {Doi}, {Dombeck}, {Eisenstein}, {Ellman}, {Elms},
  {Evans}, {Fan}, {Federwitz}, {Fiscelli}, {Friedman}, {Frieman}, {Fukugita},
  {Gillespie}, {Gunn}, {Gurbani}, {de Haas}, {Haldeman}, {Harris}, {Hayes},
  {Heckman}, {Hennessy}, {Hindsley}, {Holm}, {Holmgren}, {Huang}, {Hull},
  {Husby}, {Ichikawa}, {Ichikawa}, {Ivezi{\'c}}, {Kent}, {Kim}, {Kinney},
  {Klaene}, {Kleinman}, {Kleinman}, {Knapp}, {Korienek}, {Kron}, {Kunszt},
  {Lamb}, {Lee}, {Leger}, {Limmongkol}, {Lindenmeyer}, {Long}, {Loomis},
  {Loveday}, {Lucinio}, {Lupton}, {MacKinnon}, {Mannery}, {Mantsch}, {Margon},
  {McGehee}, {McKay}, {Meiksin}, {Merelli}, {Monet}, {Munn}, {Narayanan},
  {Nash}, {Neilsen}, {Neswold}, {Newberg}, {Nichol}, {Nicinski}, {Nonino},
  {Okada}, {Okamura}, {Ostriker}, {Owen}, {Pauls}, {Peoples}, {Peterson},
  {Petravick}, {Pier}, {Pope}, {Pordes}, {Prosapio}, {Rechenmacher}, {Quinn},
  {Richards}, {Richmond}, {Rivetta}, {Rockosi}, {Ruthmansdorfer}, {Sandford},
  {Schlegel}, {Schneider}, {Sekiguchi}, {Sergey}, {Shimasaku}, {Siegmund},
  {Smee}, {Smith}, {Snedden}, {Stone}, {Stoughton}, {Strauss}, {Stubbs},
  {SubbaRao}, {Szalay}, {Szapudi}, {Szokoly}, {Thakar}, {Tremonti}, {Tucker},
  {Uomoto}, {Vanden Berk}, {Vogeley}, {Waddell}, {Wang}, {Watanabe},
  {Weinberg}, {Yanny}, {Yasuda}, \& {SDSS Collaboration}}]{york2000}
{York}, D.~G., {Adelman}, J., {Anderson}, Jr., J.~E., {et~al.} 2000, \aj, 120,
  1579, \dodoi{10.1086/301513}

\bibitem[{{Yuan} \& {Kewley}(2009)}]{yuan2009}
{Yuan}, T.~T., \& {Kewley}, L.~J. 2009, \apjl, 699, L161,
  \dodoi{10.1088/0004-637X/699/2/L161}

\bibitem[{{Zahedy} {et~al.}(2019){Zahedy}, {Chen}, {Johnson}, {Pierce},
  {Rauch}, {Huang}, {Weiner}, \& {Gauthier}}]{zahedy2019}
{Zahedy}, F.~S., {Chen}, H.-W., {Johnson}, S.~D., {et~al.} 2019, \mnras, 484,
  2257, \dodoi{10.1093/mnras/sty3482}

\bibitem[{{Zahedy} {et~al.}(2021){Zahedy}, {Chen}, {Cooper}, {Boettcher},
  {Johnson}, {Rudie}, {Chen}, {Cantalupo}, {Cooksey}, {Faucher-Gigu{\`e}re},
  {Greene}, {Lopez}, {Mulchaey}, {Penton}, {Petitjean}, {Putman}, {Rafelski},
  {Rauch}, {Schaye}, {Simcoe}, \& {Walth}}]{zahedy2021}
{Zahedy}, F.~S., {Chen}, H.-W., {Cooper}, T.~M., {et~al.} 2021, \mnras,
  \dodoi{10.1093/mnras/stab1661}

\bibitem[{{Zahid} {et~al.}(2014){Zahid}, {Kashino}, {Silverman}, {Kewley},
  {Daddi}, {Renzini}, {Rodighiero}, {Nagao}, {Arimoto}, {Sanders},
  {Kartaltepe}, {Lilly}, {Maier}, {Geller}, {Capak}, {Carollo}, {Chu},
  {Hasinger}, {Ilbert}, {Kajisawa}, {Koekemoer}, {Kovac{\#728}}, {Le
  F{\`e}vre}, {Masters}, {McCracken}, {Onodera}, {Scoville}, {Strazzullo},
  {Sugiyama}, {Taniguchi}, \& {The COSMOS Team}}]{zahid2014}
{Zahid}, H.~J., {Kashino}, D., {Silverman}, J.~D., {et~al.} 2014, \apj, 792,
  75, \dodoi{10.1088/0004-637X/792/1/75}

\end{thebibliography}
\bibliographystyle{aasjournal}

\end{document}